\documentclass[aoas]{imsart}
\let\singlespacing\relax
\let\doublespacing\relax
\RequirePackage[OT1]{fontenc}
\RequirePackage{amsthm,amsmath}
\RequirePackage[colorlinks,citecolor=blue,urlcolor=blue]{hyperref}
\usepackage[utf8]{inputenc}
\DeclareUnicodeCharacter{FFFD}{}

\usepackage[normalem]{ulem}
\usepackage{array}
\usepackage[table]{xcolor}
\usepackage{amsmath}
\usepackage{amsfonts,mathtools}
\usepackage{graphicx}
\usepackage{epstopdf,epsfig}
\usepackage{multirow}
\usepackage{setspace}
\usepackage{natbib}
\usepackage{hyperref}
\usepackage{comment}
\usepackage[parfill]{parskip}

\newtheorem{proposition}{Proposition}

\newcommand{\bay}[1]{\textcolor{red}{#1}}
\definecolor{light-gray}{gray}{0.9}

\begin{document}
\begin{frontmatter}

% "Title of the paper"
\title{Modeling the social media relationships of Irish politicians using a generalized latent space stochastic blockmodel}
\runtitle{Modeling social media relationships of Irish politicians}

% indicate corresponding author with \corref{}
% \author{\fnms{John} \snm{Smith}\corref{}\ead[label=e1]{smith@foo.com}\thanksref{t1}}
% \thankstext{t1}{Thanks to somebody} 
% \address{line 1\\ line 2\\ printead{e1}}
% \affiliation{Some University}

\begin{aug}
\author{\fnms{Tin Lok James} \snm{Ng} \corref{}},
\author{\fnms{Thomas Brendan} \snm{Murphy}},
\author{\fnms{Ted} \snm{Westling}},
\author{\fnms{Tyler H.} \snm{McCormick}},
\author{\fnms{Bailey} \snm{Fosdick}}

\begin{comment}
\author[A]{\fnms{Tin Lok James} \snm{Ng} \corref{}\ead[label=e1]{jamesng@uow.edu.au}},
\author[B]{\fnms{Thomas Brendan} \snm{Murphy} \ead[label=e2]{brendan.murphy@ucd.ie}},
\author[C]{\fnms{Ted} \snm{Westling} 
\ead[label=e3]{twestling@umass.edu}
},
\author[D]{\fnms{Tyler H.} \snm{McCormick} 
\ead[label=e4]{tylermc@uw.edu}
},
\author[D]{\fnms{Bailey} \snm{Fosdick} \ead[label=e5]{bailey.fosdick@colostate.edu}
}

\address[A]{School of Mathematics and Applied Statistics,\\
University of Wollongong,\\ 
Australia\\ \printead{e1}}
\address[B]{
School of Mathematics \& Statistics, \\
University College Dublin, \\
Ireland\\
\printead{e2}}
\address[C]{Department of Mathematics and Statistics,\\
University of Massachusetts Amherst,\\
Amherst,\\
Massachusetts\\
\printead{e3}}
\address[D]{Deprartment of Statistics, \\
University of Washington,\\
Seattle,\\
Washington\\
\printead{e4}}
\address[E]{Department of Statistics,\\
Colorado State University,\\
Fort Collins,\\
Colorado\\
\printead{e5}}
\end{comment}
\end{aug}

\runauthor{T. L. J. Ng et al.}

\begin{abstract}
D\'{a}il \'{E}ireann is the principal chamber of the Irish parliament. The 31st D\'{a}il was in session from March 11th, 2011 to February 6th, 2016. Many of the members of the D\'{a}il were active on social media and many were Twitter users who followed other members of the D\'{a}il. The pattern of Twitter following amongst these politicians provides insights into political alignment within the D\'{a}il. We propose a new model, called the \textit{generalized latent space stochastic blockmodel}, which extends and generalizes both the latent space model and the stochastic blockmodel to study social media connections between members of the D\'{a}il. The probability of an edge between two nodes in a network depends on their respective class labels, as well as sender and receiver effects and latent positions in an unobserved latent space. The proposed model is capable of representing transitivity, clustering, as well as disassortative mixing. A Bayesian method with Markov chain Monte Carlo sampling is proposed for estimation of model parameters. Model selection is performed using the WAIC criterion and models of different number of classes or dimensions of latent space are compared. We use the model to study Twitter following relationships of members of the D\'{a}il and interpret structure found in these relationships. We find that the following relationships amongst politicians is mainly driven by past and present political party membership. We also find that the modeling outputs are informative when studying voting within the D\'{a}il. 
\end{abstract}

%\begin{keyword}[class=MSC]
%\kwd[Primary ]{}
%\kwd{}
%\kwd[; secondary ]{}
%\end{keyword}

%\begin{keyword}
%\kwd{}
%\kwd{}
%\end{keyword}

\end{frontmatter}

\section{Introduction}
Networks are widely used to represent relational data. In social network data, vertices typically represent individual nodes while edges represent ties or relationships between nodes. For example, on the Twitter social medium platform, the nodes are users and the following of users can be represented by edges. Twitter is widely used by many public figures, including politicians. In this work, we study the following connections of Irish politicians on Twitter to see if the links reveal insights about alignment of the politicians in the parliament. 

A number of network models have been proposed in recent years, including latent space models  \citep{hoff02,handcock07}, stochastic blockmodels \citep{holland83,wang87,snijders97,nowicki01}, overlapping stochastic blockmodels \citep{latouche11,latouche14}  and mixed membership stochastic blockmodels \citep{airoldi06,airoldi08}; \cite{townshend12} contains a detailed review of network models. 

The stochastic blockmodel assumes that each node belongs to one latent class. Conditional on latent class assignments, the probability of an edge between two nodes depends only on the latent classes to which the nodes belong. These models are able to represent both assortative mixing, indicated by within-block probabilities being larger than between-block probabilities, and disassortative mixing, indicated by the reverse. \cite{peng16} developed the degree-corrected stochastic blockmodel which captures the degree heterogeneity of a network. However, the degree-corrected stochastic blockmodel lacks the flexibility to capture local sturcture in a network.

The latent space model \citep{hoff02} assigns each node to a latent position in a Euclidean space. The probability of an edge between two nodes in the latent space is the logistic transformation of a linear function of the Euclidean distance between their respective latent positions and their covariates. This model takes advantage of the properties of Euclidean space to naturally represent transitivity and reciprocity. The latent position cluster model \citep{handcock07} extends the latent space model by accounting for clustering. In this model, the latent positions of nodes are drawn from a finite mixture of multivariate normal distributions. \cite{krivitsky09} further extended the latent position cluster model by allowing node-specific random effects. 

While the latent space model and its extensions are able to capture transitivity, reciprocity, and homophily, they are unable to represent disassortative mixing -- the phenomena of nodes being more likely to have ties with nodes which are different from them. Social networks tend to be assortative when patterns of friendships are strongly affected by race, age, and language; however, disassortative mixing is more common in biological networks. %\textcolor{red}{do we see evidence of disassortative mixing in the data? if so maybe add a sentence briefly desribing here?}

We propose the generalized latent space stochastic blockmodel, which addresses the issues of stochastic blockmodels as well as latent space models. The resulting model can be viewed as a generalization of stochastic blockmodels as well as latent space models. The model generalizes the latent space stochastic blockmodel recently developed in \cite{fosdick16} on several fronts. The model permits directed edges and relaxes the assortative mixing assumption made in \cite{fosdick16}. Both of these extensions are essential for effectively modeling the Irish politician Twitter following data. More importantly, the model incorporates sender and receiver effects for modelling between block connections, which greatly enhance the flexibility of the model. The proposed model can also be seen as an extension of the degree corrected stochastic blockmodel \citep{peng16} by modelling the within block connections using latent spaces.

\cite{schweinberger14} introduced the general concept of local dependence which allows separate modelling of within class and between class connection probabilities. Assortative mixing of nodes is enforced by constraining between class connection probabilities to be smaller than within class probabilities. In contrast, by allowing both assortative and disassortative mixing, the proposed model is capable of representing a wide range of real networks.

In Section~\ref{se:dail} we introduce a data set that contains the social media connections between Irish politicians on the Twitter social media platform. In Section~\ref{se:model} we describe the generalized latent space stochastic blockmodel (GLSSBM). In Section~\ref{se:mcmc} we describe model estimation based on Monte Carlo Markov Chain (MCMC) sampling. In Section~\ref{se:selection} we propose a method for model comparison so that the number of classes and latent space dimension can be selected. In Section~\ref{se:simulation} we perform simulation studies to investigate the performance of the proposed model and the MCMC algorithm. In Section~\ref{se:application} we use the generalized latent space stochastic blockmodel to investigate the Twitter following relationships between the Irish politicians. We discuss the structure that is revealed by the model and its connection to the politicians' political alignment within the parliament. In Section~\ref{se:discussion} we discuss the model and possible extensions. 

\section{Irish Politicians Twitter Data}
\label{se:dail}

The Irish legislature is called the Oireachtas, and it consists of the President of Ireland and a parliament with two chambers: Seanad \'{E}ireann and D\'{a}il \'{E}ireann. D\'{a}il \'{E}ireann is the principal chamber of the Irish parliament, and it is directly elected from constituencies that elect either three, four or five seats (members). Each seat represents between 20,000 and 30,000 Irish citizens. Further details of the Irish political system can be found in \cite{coakley18}.

The 31st Irish D\'{a}il (March 11th, 2011 - February 6th, 2016) had 166 members who were affiliated with nine political parties or were independent. Of these, 147 members had accounts on the Twitter social media platform (Table~\ref{table:dail}). The government consisted of a coalition between Fine Gael and Labour and the opposition had a number of parties including Fianna F\'{a}il, Sinn Fein, as well as a number of small parties and independent politicians. 
\begin{table}[ht]
    \caption{Political Parties in the 31st Irish D\'{a}il on January 1st, 2016. The number of members of each party and the number Twitter users are also shown. On this date, Fine Gael and Labour were in a coalition government.
    %\textcolor{red}{(Can we highlight groups that typically are in coalition together?  As well as those that represent the government.  This information is buried in the text but it would be great to have it in a place that is easy to find.)}
    }
    \label{table:dail}
\smallskip
\centering
{ \begin{tabular}{lrr}
  \hline\hline
Party & Members & Twitter \\ 
  \hline
Fianna F\'{a}il &  21&  17 \\ 
  Fine Gael &  66& 61 \\ 
  Independent &  18& 17 \\ 
  Labour &  34&  29 \\ 
  People Before Profit Alliance & 1 &  1 \\ 
  Renua Ireland & 3& 2 \\ 
  Sinn F\'{e}in & 14&  12 \\ 
  Social Democrats & 3&   3 \\ 
  Socialist Party &  3& 3 \\ 
  United Left &   2& 2 \\ 
  \hline
   All & 166 & 147 \\
   \hline\hline
\end{tabular}}
\end{table}

\begin{table}[ht]
    \caption{Average in-degree and average out-degree for each political party.
    }
    \label{table:dail_degree}
\smallskip
\centering
{ \begin{tabular}{lrr}
  \hline\hline
Party & Average in-degree & Average out-degree \\ 
  \hline
Fianna F\'{a}il &  34.3 &  28.3 \\ 
  Fine Gael &  57.6 & 56.7 \\ 
  Independent &  34.2 & 30.1 \\ 
  Labour &  47.5 &  49.3 \\ 
  People Before Profit Alliance & 17.0 &  32.0 \\ 
  Renua Ireland & 68.0 & 58.5 \\ 
  Sinn F\'{e}in & 17.7 &  20.6 \\ 
  Social Democrats & 22.7 &   53.3 \\ 
  Socialist Party &  11.3 & 17.0 \\ 
  United Left &   11.0 & 26.5 \\ 
   \hline\hline
\end{tabular}}
\end{table}

\begin{figure}[ht]
\begin{center}
\includegraphics[width=0.7\textwidth]{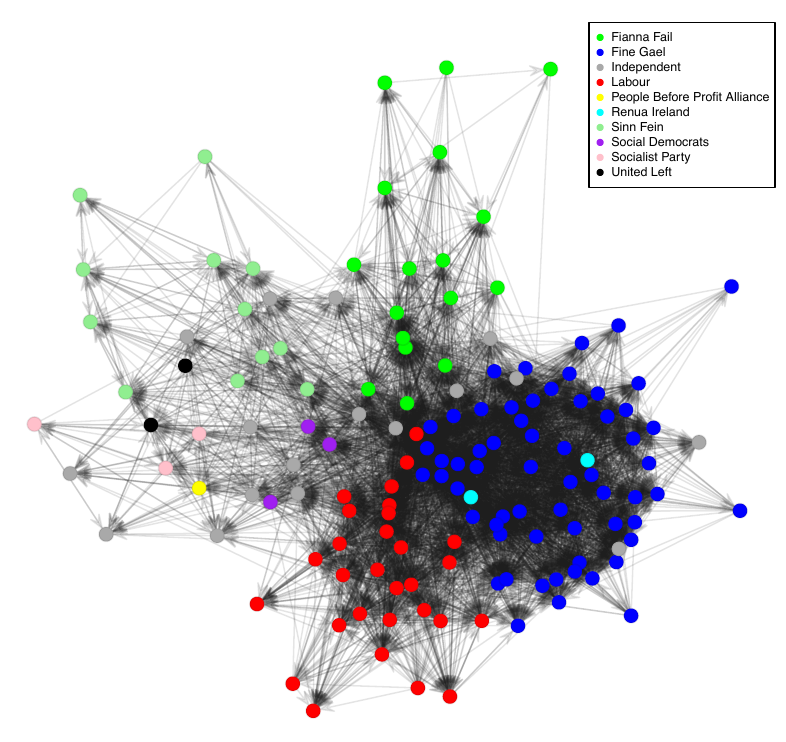}
\end{center}
\caption{A force directed layout of the Irish politicians Twitter follower network. The nodes are colored by political affiliation of each member of the parliament with a Twitter account.}
\label{fi:daillayout}
\end{figure}

A directed network consisting of the following relationship of the 147 accounts was constructed, where there is an edge from member $i$ to member $j$ if member $i$ follows member $j$. A plot of the network is given in Figure~\ref{fi:daillayout}, where the nodes are colored according to their political party. Table~\ref{table:dail_degree} shows the average in-degree and average out-degree for members in each political party.

The plot in Figure~\ref{fi:daillayout} shows that nodes within the same party are clustered together. The Fruchterman Reingold force-directed layout used in the plot puts nodes closer together if they are connected. Thus, there is an indication that political party affiliation is related to Twitter following.

Data were collected on the voting within the D\'{a}il from January 1st, 2016 to February 6th, 2016. In this time period, a total of 23 votes took place. Each member of the D\'{a}il was recorded as voting T\'{a} (Yes), N\'{i}l (No) or Absent/Abstain for each vote. The period was characterized by having a large number of Absent/Abstain, with 30.7--71.7\% abstaining in the votes, which can be explained by a few factors. Firstly, the political parties have an informal agreement that on some votes both the government and opposition will have the same number of members absent so that the outcome of the vote isn't altered.  Secondly, the term of the D\'{a}il is five years, so the politicians knew that the election would be called before March 11th, 2016; the date of the election is chosen by the Taoiseach (Head of Government). Thus, many politicians were campaigning locally in their constituencies in anticipation of the upcoming election.

We are interested in exploring the structure of the Twitter following network for the members of the Irish parliament. In particular, we want to determine if there are subgroups in the network with similar connectivity patterns, to determine if political party is related to the structure in the network, and to determine whether the voting in the D\'{a}il is related to the structure in the network. These questions are particularly interesting in a parliament with such a large number of independent members, because there is less prior information available on how these members interact with each other and with the members who are affiliated with political parties. 

It is worth noting that Fianna F\'{a}il and Fine Gael are traditionally the two main parties in Ireland, but Fianna F\'{a}il had their worst election performance ever in the election of the 31st D\'{a}il. Members of these parties tend to take opposite positions on many matters. Fine Gael and Labour have been in coalition on a number of occasions, with Labour being the minor party of the coalition, but Fianna F\'{a}il and Labour were in coalition from December 14th, 1992 to December 15th, 1994. Fine Gael and Labour are closely aligned on many policies but take different stances on other policies. \cite{hansen09} used a Bayesian ideal point model to estimate the location of the parties relative to each other and the results showed a government-opposition divide that explained the location of parties relative to each other.

\cite{hansen10} found that Fianna F\'{a}il and Fine Gael tend to vote in opposite ways in the parliament, as do Fianna F\'{a}il and Labour, whereas Fine Gael and Labour tend to vote together.  \cite{hansen10} also found that minor party members tend to vote against the government unless they are part of the government coalition, whereas independent members tend to vote en bloc either in support of or in opposition to the government, depending on the context. 

\cite{bolleyer09} provide a detailed description of the role of independents in the Irish context from the electoral stage through to their role in parliament and in government. \cite{weeks09} studied the role of independent candidates in Irish politics and provided a typology of independent politicians (not just those elected). His typology had six categories of independents, and those most relevant in this data are {\em national single-issue independents}, who are concerned with some ideological national issue, {\em community independents}, who are entirely focused on representing their local community with little concern in national issues, and {\em apostate independents}, who flit between party and independent status (some of these either resigned or were expelled from a party or failed to secure a nomination to be a candidate for the party). Many of the independent members of the 31st D\'{a}il can be categorized into these categories. 

\cite{farrell15} discuss the role of political party on voting within the D\'{a}il, arguing that Irish politicians have historically shown strong party discipline in voting. One reason for this is that the political parties impose high penalties for those who do not follow the party line. However, the Irish electoral system is very candidate-focused, with candidates from the same party competing for election in a constituency \citep[eg.][]{gormley08,gormley08a}. However, historically, very few candidates have rebelled against the party when voting in the D\'{a}il.

\section{Model}
\label{se:model}

\subsection{Generalized Latent Space Stochastic Blockmodel}
The network being modeled is a binary, directed network, though the model developed herein can easily be adjusted to undirected networks. We let $N$ denote the number of nodes (politicians) in the network and denote the $N\times N$ adjacency matrix by ${\mathbb Y}_N$. We let $y_{ij}=1$ if node $i$ follows node $j$ and $y_{ij}=0$ otherwise. 

We partition nodes into $K$ latent classes and denote the class assignment by ${\boldsymbol \gamma} = (\gamma_{1}, \gamma_{2}, \ldots, \gamma_{N}) $. We assume that the class assignments of the nodes are independent and identically distributed. That is,
\begin{equation*}
\gamma_i|\boldsymbol{\pi} \sim \mbox{multinomial}( 1; \pi_{1}, \ldots, \pi_{K} ),
\end{equation*}
where $\pi_k$ is the probability of a node being assigned to class $k$. We also assign each node a latent position in the Euclidean space ${\mathbb R}^{d}$, the collection of which is denoted by $ \textbf{Z} = (Z_{1}, Z_{2}, \ldots, Z_{N}) $. We further assume that the latent position $Z_{i}$ is drawn from a multivariate normal distribution $ Z_{i} \sim  \mbox{MVN}( 0, (\sigma^{z})^2 I_{d} ) $, 
where $\boldsymbol{\sigma} = ( \sigma_{1}, \ldots, \sigma_{K} )$ are the standard deviations of the latent positions within each class.

If two nodes belong to the same class \textit{k}, we assume that the probability there is an edge between them depends only on their latent positions, and model the connection probability as
\begin{equation*}
  \mbox{ logit } P( Y_{ij} = 1 \mid \gamma_{i} = \gamma_{j} = k, Z_i, Z_j ) = \beta_{k} - \| Z_{i} - Z_{j} \| .
\end{equation*}
In particular, we assume that relationships within blocks tend to be reciprocal, that is, the probability of link from $i$ to $j$ is the same as the probability of a link from $j$ to $i$ if both $i$ and $j$ belong to the same class. If two nodes have different class labels, we assume that the presence of an edge depends both on their respective class labels and their sender and receiver effects. That is,
$$Y_{ij}| \gamma_i=k, \gamma_j = l, s_i, r_j, k \ne l \sim \mbox{Bern}(\mbox{logit}^{-1}(\eta_{kl}+ s_i + r_j) ) $$
where $\mathbf{s} = (s_1, \ldots, s_N)$ and $\mathbf{r} = (r_1, \ldots, r_N)$ are the sender and receiver effects, and $\boldsymbol{\eta} = (\eta_{kl})_{k,l}$ is the block-to-block interaction matrix. By assumption, the diagonal entries of $\boldsymbol{\eta}$ are all zero.

Hence, conditional on class labels, latent positions and sender and receiver effects, the likelihood function is:
\begin{equation}
\label{eq:like}
     \mbox{P}(\mathbb{Y}_N | \boldsymbol{\gamma}, \mathbf{Z}, \boldsymbol{\theta}, \boldsymbol{\beta}) = \prod_{i \ne j} (p_{ij})^{Y_{ij}} (1-p_{ij})^{1-Y_{ij}},
\end{equation}
where $p_{ij} = \mbox{logit}^{-1}(\beta_k - ||Z_i - Z_j||)$ if $\gamma_i = \gamma_j = k$, and $p_{ij} = \mbox{logit}^{-1}(\eta_{kl} + s_i + r_j)$ if $\gamma_i = k, \gamma_j = l$ and $k \ne l$.

\subsection{Model identifiability}
We investigate the identifiability of the proposed model in this section. We follow the same approach as in \cite{peng16} to group the parameters as
$$\boldsymbol{\theta} := (\eta_{12}, \ldots, \eta_{1K}, \ldots, \eta_{K,K-1}, s_1, \ldots, s_N, r_1, \ldots, r_N)^{T}.$$ 
We define the design matrix $X(\boldsymbol{\gamma})$, which depends on $\boldsymbol{\gamma}$, associated with associated with $\boldsymbol{\theta}$ as
$$Y_{ij} | \gamma_i = k, \gamma_j = l \sim \mbox{Bern}\big(\mbox{logit}^{-1}( \mathbf{x}_{ij}^{T} \boldsymbol{\theta} )\big) ,$$
for every pair $(i,j)$ such that $\gamma_i \ne \gamma_j$, and where each $\mathbf{x}_{ij}$ corresponds to a row in $X$. In particular, for $\gamma_i \ne \gamma_j$, we have
$$\mathbf{x}_{ij}^{T} \boldsymbol{\theta} = \eta_{\gamma_i \gamma_j} + s_i + r_j .$$

As shown in the following proposition, the generalized latent space stochastic blockmodel developed above is not identifiable, with $2K$ redundant parameters.

\begin{proposition}
The design matrix $X(\boldsymbol{\gamma})$ has 2$K$ linearly dependent columns.
\end{proposition}
\begin{proof}
The proof follows similar lines as in \cite{peng16}. We can split each row $\mathbf{x}_{ij}$ according to its $\boldsymbol{\eta}, \mathbf{r}, \mathbf{s}$ entries, $\mathbf{x}_{ij} = (\mathbf{b}_{ij}, \mathbf{c}_i, \mathbf{d}_j)$, where
$$ b_{ij,kl} = I(\gamma_i=k, \gamma_j=l), \quad k \ne l,k,l=1,\ldots,N$$
$$ c_{i,v} = I(i=v), \quad v=1,\ldots, N, $$
$$ d_{j,v} = I(j=v), \quad v=1, \ldots, N.$$
For an arbitrary block $k$, we have that
$$ \sum_{l=1, l \ne k}^{K} b_{ij,kl} = \sum_{v: \gamma_v = k} c_{i,v} .$$
Similarly, for an arbitrary block $l$, we have
$$ \sum_{k=1, k \ne l}^{K} b_{ij,kl} = \sum_{v: \gamma_v = l} d_{j,v} .$$
Since the above two equalities hold for each $k=1,\ldots,K$ and $l=1,\ldots,K$, the design matrix $X(\boldsymbol{\gamma})$ has $2K$ linearly dependent columns.
\end{proof}

To ensure identifiability of the model, we impose the following $2K$ constraints on the parameters:
$$ \sum_{i: \gamma_i = k} s_i = 0, \sum_{i: \gamma_i = k} r_i = 0,$$
for $k = 1, \ldots, K$. That is, the sum of the sender and receiver effects in each block are set to zero. In an MCMC run, these constraints can be enforced at the end of each iteration by making the following adjustments to the block-to-block interaction probabilities $\boldsymbol{\eta}$ and sender and receiver effects, $\mathbf{s}$ and $\mathbf{r}$. 

Suppose $\boldsymbol{\eta}, \mathbf{s}, \mathbf{r}$ are the sampled parameters at an iteration of an MCMC run, for $k=1,\ldots,K$, we define $$ S_k := \sum_{i: \gamma_i = k} s_i, \quad R_k = \sum_{i: \gamma_i = k} r_i .$$

Let $n_k$ be the number of nodes in block $k$, for $k=1,\ldots,K$. For each pair of blocks $k \ne l$, we set
$$ \eta'_{kl} = \eta_{kl} + \frac{1}{n_k} S_k + \frac{1}{n_l} R_l ,$$
and for each node $i$ with $\gamma_i = k$, we set
$$ s_i' = s_i - \frac{1}{n_k} S_k,$$
$$ r_i' = r_i - \frac{1}{n_k} R_k.$$

Furthermore, the likelihood function is invariant to reflections, rotations, and translations of the latent spaces and permutation of class labels. As in the case of latent position cluster model \citep{handcock07}, the non-identifiability issues can be resolved by Procrustes transformation \citep{sibson79} and post-processing the MCMC output. We first fix a reference set of latent positions for the nodes, which can be obtained by a  standard method of multidimensional scaling to one minus the adjacency matrix of the network. We then post-process the MCMC output by applying the relabelling algorithm \citep{celeux00} to solve the label switching problem. We also post-process the MCMC output by perform a Procrustes transformation with translation and rotation to minimize the distance between simulated and reference latent positions.

\section{Bayesian Estimation}
\label{se:mcmc}
\subsection{Prior Specification}
We fit the model specified in Section~\ref{se:model} in a Bayesian paradigm using MCMC sampling. We assign prior distributions for the parameters $ \boldsymbol{ \pi } $,  and $ \boldsymbol{ \beta } $ as follows:
\begin{align*}
  \boldsymbol{\pi} &\sim \mbox{Dirichlet}(T,\ldots,T), \\
   \beta_{k} &\sim \mbox{N}(0, ( \sigma^{\beta} )^2),
\end{align*}
where $1  \le k \le K$. The parameter vector $\boldsymbol{\theta}$ which contains block-to-block interaction parameters $\boldsymbol{\eta}$, and sender and receiver effects, $\mathbf{s}$ and $\mathbf{r}$, is assigned a multivariate normal prior:
$$ \boldsymbol{\theta} \sim \mbox{MVN}\big(0_{2N+K^2-K}, (\sigma^{\theta})^{2} I_{2N+K^2-K}\big) ,$$
where $0_{2N+K^2-K}$ is a vector of $0$s, and $I_{2N+K^2-K}$ is the identity matrix. Recall from Section \ref{se:model} that the latent positions $Z_i \sim \mbox{MVN}(0, (\sigma^{z})^{2} I_d)$.

As discussed in \cite{nowicki01}, small values of $T$ tend to favor unequal class sizes. For the Twitter follower data application, we set $T = 10$ to discourage very small classes that are almost empty. The prior standard deviations are set as $\sigma^{\beta} = 5, \sigma^z = 3, \sigma^{\theta} = 5$. The choice of the hyper-parameters lead to good predictive performance as demonstrated in simulation studies and real application.

\subsection{Full conditional distributions}
\label{subsec:full_cond}
With the prior assumptions given above, the full conditional posterior distributions can be expressed as:
\begin{equation}
   \label{eq:pi_pos}
   [{\boldsymbol\pi}\mid others] \sim \mbox{Dirichlet}(T+n_{1},\ldots,T+n_{c})
\end{equation}
\begin{equation}
  [\beta_{k}\mid others] \propto \bigg\{ \prod_{\gamma_i = \gamma_j = k} (p_{ij})^{Y_{ij}} (1-p_{ij})^{1-Y_{ij}} \bigg\}
   \mbox{N}(0, ( \sigma^{\beta} )^2)
\end{equation}
\begin{equation}
\label{eq:gamma_pos}
  [\gamma_i = k \mid others] \propto \pi_k \prod_{j \ne i} \bigg\{ (p_{ij})^{Y_{ij}} (1-p_{ij})^{1-Y_{ij}} \bigg\}   
  \bigg\{ (p_{ji})^{Y_{ji}} (1-p_{ji})^{1-Y_{ji}} \bigg\}  
\end{equation}
\begin{eqnarray}
\label{eq:z_pos}
    [Z_i|\gamma_i = k, others] \propto \prod_{j \ne i} \bigg\{ (p_{ij})^{Y_{ij}} (1-p_{ij})^{1-Y_{ij}} \bigg\}   
  \bigg\{ (p_{ji})^{Y_{ji}} (1-p_{ji})^{1-Y_{ji}} \bigg\} \nonumber
  \\ \mbox{MVN}\big(Z_i;0_d,(\sigma^{z})^{2}\big)
\end{eqnarray}
where $p_{ij}$ is defined in the likelihood function \eqref{eq:like}, and it is assumed that $\gamma_i = k$ on the RHS of \eqref{eq:gamma_pos} and \eqref{eq:z_pos}.

The full conditional posterior of $\boldsymbol{\theta}$ does not have a closed form expression. However, employing a data augmentation strategy by introducing additional latent variables from a Polya-Gamma distribution \citep{polson2013, peng16}, the conditional posterior of $\boldsymbol{\theta}$ can be expressed in closed form.

Define the latent variables $\boldsymbol{\omega} = (\omega_{ij})_{i \ne j: \gamma_i \ne \gamma_j} $ by
$$ \omega_{ij} | \mbox{others} \sim \mbox{PG}(1, \mathbf{x}_{ij}(\gamma)' \boldsymbol{\theta}) $$
where PG denotes the Polya-Gamma distribution. We have
$$ \boldsymbol{\theta} | \boldsymbol{\omega}, \mbox{others} \sim \mbox{MVN}(m, V) $$
where, with $\Omega = \mbox{Diag}(\omega_{ij})$ and $u_{ij} = (Y_{ij} - 1/2) \omega_{ij}^{-1} $, $\mathbf{u} = (u_{ij})_{i \ne j}$
$$ m = V X^{T} \Omega \mathbf{u} ,$$ and
$$ V = (X^{T} \Omega X + (\sigma^{\theta})^{2} I_{2N+K^2-K} )^{-1} .$$

\subsection{MCMC Algorithm}
In designing the algorithm to sample from the posterior distribution, we first note that if two nodes have difference class labels, their respective latent positions are irrelevant. Hence, the class label $\gamma_i$ and the latent position $Z_i$ of a node are updated simultaneously. On iteration $t+1$ of the algorithm, we use the following proposal to propose a new pair $(\gamma_i^{*}, Z_i^{*})$ from the existing pair $(\gamma_i^{t}, Z_i^{t})$.
\\
We let 
\begin{eqnarray}
\label{joint_pro1}
q((\gamma_i^{t}, Z_i^{t}), (\gamma_i^{*}, Z^{*}_i)) = q(\gamma_i^{t}, \gamma_i^{*}) q(Z_i^{t}, Z_i^{*}|\gamma_i^{*}) 
\end{eqnarray}
be the proposal distribution for the class label and latent position, where $q(\gamma_i^{t}, \gamma_i^{*})$ is the probability of proposing the class label $\gamma_i^{*}$ given current class label $\gamma_i^{t}$. The proposal distribution is specified in \eqref{eq:gamma_pos}. 
\\
If $\gamma_{i}^{*} = \gamma_{i}^{t}$, we let
\begin{gather}
   \label{joint_pro2}
   q( Z_{i}^{t}, Z_{i}^{*}\mid\gamma_{i}^{*} ) = \mbox{MVN}(Z_{i}^{*}; Z_{i}^{t}, \delta_z^2 I_{d}) 
\end{gather}

If $\gamma_{i}^{*} \ne \gamma_{i}^{t}$ and $\sum_{j \ne i} {\cal I}(\gamma_{j} = \gamma_{i}^{*}) > 0$, that is, if the proposed class is non-empty, we let
\begin{equation}
    \label{joint_pro3}
    q(Z_{i}^{t},Z_{i}^{*}\mid\gamma_{i}^{*}) = \mbox{MVN}(Z_{i}^{*}; \overline{Z_{\gamma^{*}}}, \delta I_{d} )
\end{equation}
where $ \overline{Z_{\gamma_{i}^{*}}} $ is the average latent position of nodes in class $ \gamma_{i}^{*} $, and we set $\delta = 1$.

If $\gamma_{i}^{*} \ne \gamma_{i}^{t}$ and $\sum_{j \ne i} {\cal I}(\gamma_{j} = \gamma_{i}^{*}) = 0$, that is, if the proposed class is empty, we let
\begin{equation}
   \label{joint_pro4}
   q(Z_{i}^{t},Z_{i}^{*}\mid\gamma^{*}) = \mbox{MVN}(Z_{i}^{*}; 0_d, (\sigma^z)^{2} I_{d}).
\end{equation}
That is, we simulate from the prior distribution for the latent position. 

Our Metropolis-within-Gibbs algorithm \citep{metropolis53} to sample from the posterior is then as follows:

\textbf{Step 1}: 

For vertices $i = 1,2, \ldots, n $, we use Metropolis-Hastings to update $(\gamma_{i}, Z_{i})$.
\begin{enumerate}
   \item  Propose $ (\gamma_{i}^{*}, Z_{i}^{*}) $ according to equations \eqref{joint_pro1} - \eqref{joint_pro4}.
   \item  With probability equal to
   \begin{equation*}
       \frac{\mbox{P}(\mathbb{Y}_N | \boldsymbol{\gamma}^{*}, \mathbf{Z}^{*}, \boldsymbol{\theta}, \boldsymbol{\beta}) q((\gamma_i^{t}, Z_i^{t}), (\gamma^{*}_i, Z_i^{*}))}{\mbox{P}(\mathbb{Y}_N | \boldsymbol{\gamma}^{t}, \mathbf{Z}^{t}, \boldsymbol{\theta}, \boldsymbol{\beta}) q((\gamma_i^{*}, Z_i^{*}), (\gamma^{t}_i, Z_i^{t}))}
   \end{equation*}

         set $(\gamma_{i}^{t+1},Z_{i}^{t+1}) = (\gamma_{i}^{*}, Z_{i}^{*})$. 
         Otherwise, set $(\gamma_{i}^{t+1},Z_{i}^{t+1}) = (\gamma_{i}^{t}, Z_{i}^{t})$.
\end{enumerate}

\textbf{Step 2}: 

For $k =1,2, \ldots, K$, we use Metropolis-Hasting to update $ \beta_{k}$.
\begin{enumerate}
   \item Propose $ \beta_{k}^{*} \sim \mbox{N}( \beta_{k}^{t}, \delta_{\beta} )$.
   \item With probability equal to 
   \begin{equation*}
       \frac{\mbox{P}(\mathbb{Y}_N | \boldsymbol{\gamma}^{*}, \mathbf{Z}^{*}, \boldsymbol{\theta}, \boldsymbol{\beta}^{*})p( \beta_{k}^{*}; \mu^{\beta}, (\sigma^{\beta})^2 )}{\mbox{P}(\mathbb{Y}_N | \boldsymbol{\gamma}^{*}, \mathbf{Z}^{*}, \boldsymbol{\theta}, \boldsymbol{\beta}^{t})p( \beta_{k}^{t}; \mu^{\beta}, (\sigma^{\beta})^2 )}
   \end{equation*}
   
       set $\beta_{k}^{t+1} = \beta_{k}^{*}$. Otherwise, set $\beta_{k}^{t+1} = \beta_{k}^{t}$.
\end{enumerate}
\textbf{Step 3}:
Update $\boldsymbol{\theta}$ using the data augmentation approach as described in Section \ref{subsec:full_cond}.

\textbf{Step 4}:

Update ${\boldsymbol\pi}$ according to \eqref{eq:pi_pos}.

\section{Model selection}
\label{se:selection}

A number of model selection criteria and algorithms have been proposed for selecting the latent space dimension and number of clusters in the latent position cluster model. The majority of methods are concerned with choosing the appropriate number of clusters. \cite{handcock07} used a BIC approximation to the conditional posterior model probabilities of the latent position cluster model, where the conditioning is on an estimate of latent positions. For each value of the number of clusters, they computed the conditional posterior model probabilities and chose the value of the number of clusters which returned the highest probability. \cite{ryan17} proposed a Bayesian model selection method for the latent position cluster model by collapsing the model to integrate out the model parameters which allows posterior inference over the number of components. 

Likewise, a few model selection criterion have been proposed for selecting the number of groups for a stochastic blockmodel. \cite{come15} proposed the integrated classification likelihood criterion. Their focus was on choosing the number of classes and maximum {\em a posteriori} estimates of the class labels for each observation with other parameters integrated out. \cite{wang17} proposed a penalized likelihood criterion for selecting the optimal number of blocks.    

In this paper, we adopt the Watanabe-Akaike information criterion (WAIC) \citep{watanabe10} to choose the number of blocks for the generalized latent space stochastic blockmodel. WAIC is a method for estimating point-wise out-of-sample prediction accuracy from a fitted Bayesian model and can be interpreted as a computationally convenient approximation to cross-validation by using posterior simulations \citep{gelman14}.

Let $ \{ (\boldsymbol{\gamma}^{t}, \textbf{Z}^{t}, \boldsymbol{\theta}^{t}, \boldsymbol{\beta}^{t}) \}_{t=1}^{S}$ denote the posterior samples, and let $V_{t=1}^{S} \log \textbf{P}( {\mathbb Y}_N \mid \boldsymbol{\gamma}^{t}, \textbf{Z}^{t}, \boldsymbol{\theta}^{t}, \boldsymbol{\beta}^{t} ) $ denote the sample variance of the log-likelihood evaluated at the posterior samples. Then the WAIC for the generalized latent space stochastic blockmodel is given by
\begin{eqnarray}
\label{eqn:waic}
  \mbox{WAIC} &=& \log \left( \frac{1}{S} \sum_{t=1}^{S} \textbf{P}( {\mathbb Y}_N \mid \boldsymbol{\gamma}^{t}, \textbf{Z}^{t}, \boldsymbol{\theta}^{t}, \boldsymbol{\beta}^{t}  ) \right) \\ && - V_{t=1}^{S} \log \textbf{P}( {\mathbb Y}_N \mid \boldsymbol{\gamma}^{t}, \textbf{Z}^{t}, \boldsymbol{\theta}^{t}, \boldsymbol{\beta}^{t} ) \\
    &=& \mbox{(prediction accuracy)} - \mbox{(penalty)}   \mbox{   .} \nonumber
\end{eqnarray}
The first term on the right-hand side of equation \eqref{eqn:waic} measures the model fit, while the second term penalizes model complexity; models with higher WAIC value are preferred.

\section{Simulation Studies}
\label{se:simulation}
In this section we perform simulation studies to investigate the predictive performance of GLSSBM and whether the MCMC algorithm is able to recover the model parameters.
\subsection{Predictive performance}
We conduct a simulation study to compare the predictive performance of the GLSSBM with the degree-corrected stochastic blockmodel and the latent position cluster model. Three networks with 100 nodes each were generated according to a 3-block GLSSBM, 3-block degree-corrected stochastic blockmodel, and 3-cluster latent position cluster model, respectively. The model parameters were generated by appropriate prior distributions for each of the 3 cases. For each of the simulated network, 5\% of the dyads were randomly removed and the three models were fitted to the data, and their predictions on the missing dyads were obtained. 

\begin{figure}[htbp]
\begin{align*}
\includegraphics[width=0.45\textwidth]{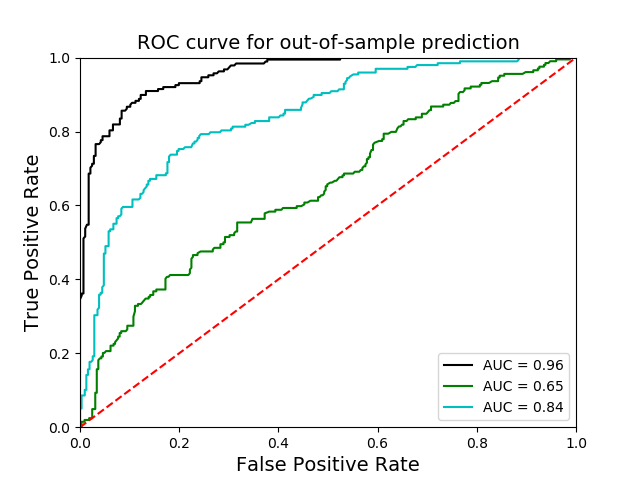} &
\includegraphics[width=0.45\textwidth]{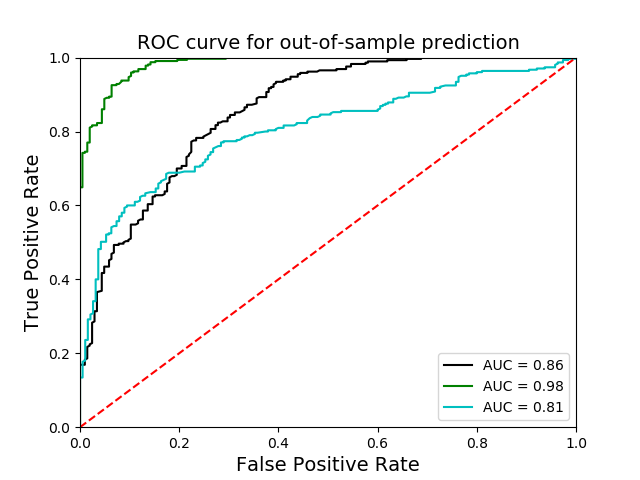} & \\
\includegraphics[width=0.45\textwidth]{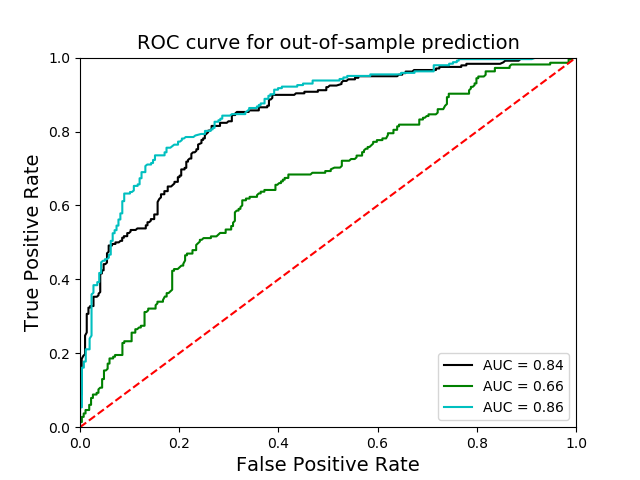}
\end{align*}
\caption{ROC curve for out-of-sample predictions for a 3-block degree-corrected stochastic blockmodel (blue), a 3-cluster latent position cluster model (green) and a 3-block GLSSBM (black). Top left: network simulated from GLSSBM. Top right: network simulated from latent position cluster model. Bottom: network simulated from degree-corrected stochastic blockmodel.} \label{fig:ROC_sim}
\end{figure}

Figure~\ref{fig:ROC_sim} shows the ROC curves for out-of-sample predictions for the three cases, respectively. We observe that the GLSSBM achieves superior predictive performance compared to the degree-corrected stochastic blockmodel when the true network is generated from the latent position cluster model. Similarly, the GLSSBM performs substantially in predictions better than the latent position cluster model when the true network is drawn from the degree-corrected stochastic block model.

\subsection{Recovery of model parameters}
\begin{figure}[htbp]
\begin{align*}
\includegraphics[width=0.48\textwidth]{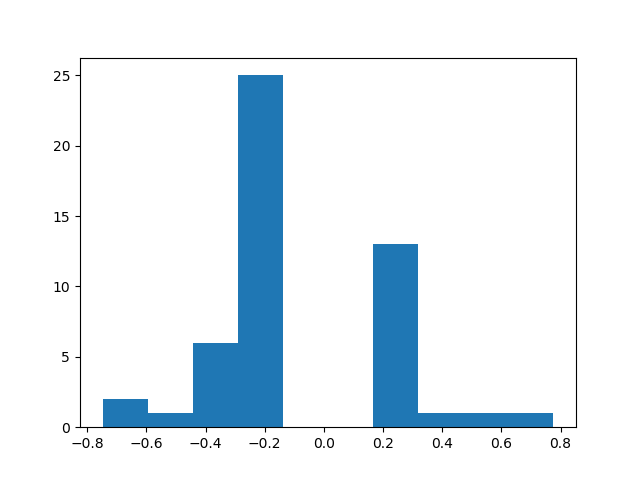} &
\includegraphics[width=0.48\textwidth]{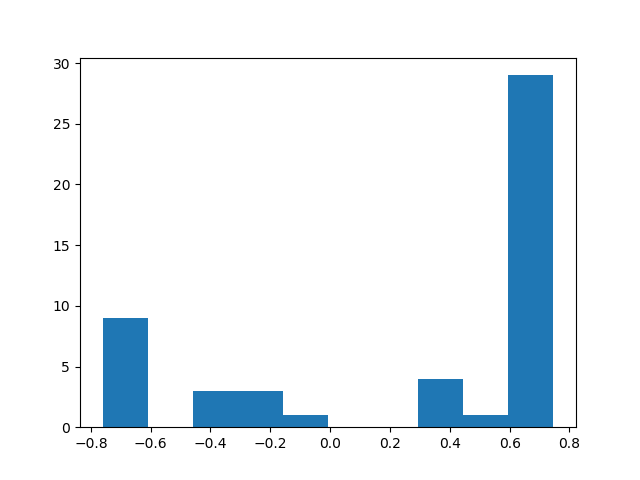} & \\
\includegraphics[width=0.48\textwidth]{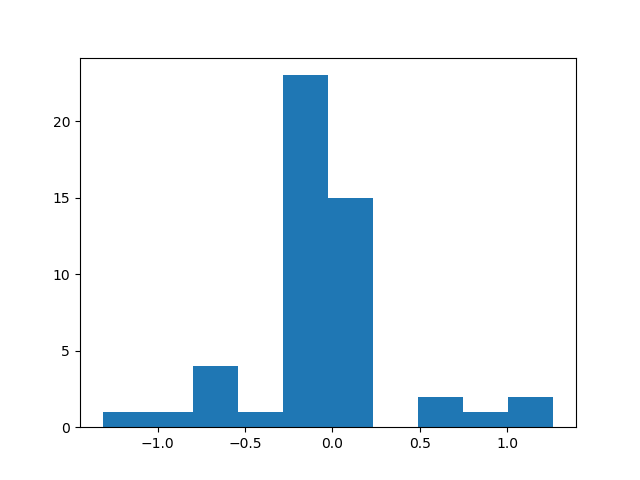}&
\includegraphics[width=0.48\textwidth]{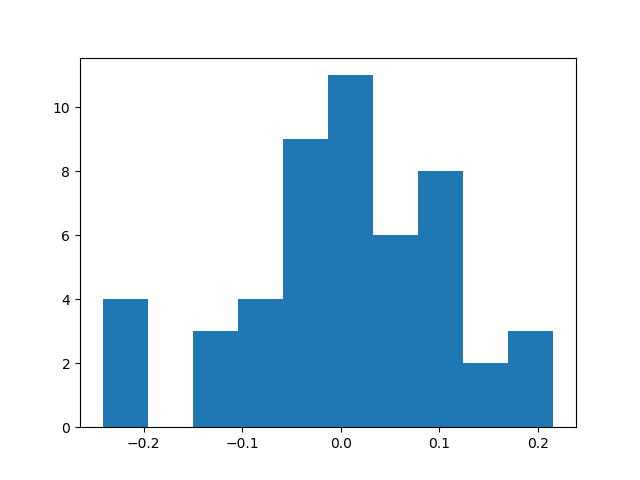}
\end{align*}
\caption{Histogram of the difference between the posterior mean estimate and the true parameters. Top left: $\hat{\eta}_{12} - \eta_{12}$. Top right: $\hat{\eta}_{21} - \eta_{21}$. Bottom left: $\hat{\beta}_{1} - \beta_1$. Bottom right: $\hat{\beta}_2 - \beta_2$.} \label{fig:parameters_recovery}
\end{figure}

We investigate whether the block-level parameters and block labels can be recovered using the proposed MCMC algorithm. 50 networks with 100 nodes each were simulated from the 2-block GLSSBM where the parameters were drawn using the prior distributions described in Section~\ref{se:mcmc}. The simulated networks were fitted using the MCMC algorithm and the posterior samples were obtained. Figure~\ref{fig:parameters_recovery} shows the distribution of the different between the posterior mean and the true value of the block-level parameters. We observe that the posterior mean estimate of the parameters and the true parameters are close. The average mis-classification rate for the block labels is 0.4\%.

\section{Irish Politician Twitter Data Structure}
\label{se:application}

We fit the generalized latent space stochastic blockmodel with varying number of blocks and dimensions to the Irish politicians Twitter follower data. Parameter estimation for the stochastic blockmodels is performed using Gibbs sampling \citep{nowicki01}. We apply the MCMC algorithm in \citep{hoff02} to fit the latent space models. We perform model selection using WAIC where the results are given in Table~\ref{table:waic}. The GLSSBM with 6 blocks is the optimal model based on WAIC. For the optimal model, the MCMC algorithm takes on average 4.1 seconds per iteration. We find that further increasing the dimensions of the latent space does not lead to increase in the WAIC. For the same number of blocks, models with $d = 2$ consistently outperforms models with $d = 3$ in terms of WAIC.

\begin{table}[htbp]
    \caption{Model selection using WAIC. The model with the highest WAIC is the most preferred (marked in bold).}
    \label{table:waic}
    \begin{center}
  \begin{tabular}{lllll}\hline\hline
     \textbf{Model} & \textbf{Blocks/Clusters} & \textbf{Pred} &\textbf{Penalty} & \textbf{WAIC}\\ \hline
     GLSSBM & 2 & -7370 & 270 & -7640 \\
     GLSSBM & 3 & -7214 & 281 & -7495 \\
     GLSSBM & 4 & -7056 & 299 & -7355 \\
     GLSSBM & 5 & -6815 & 324 & -7139 \\
     GLSSBM & {\bf 6} & {\bf -6570} & {\bf 352} & {\bf -6922} \\
     GLSSBM & 7 & -6680 & 389  & -7069 \\
     DCSBM & 1 & -9429  & 142  & -9571 \\
     DCSBM & 2 & -8290  & 184  & -8474  \\
     DCSBM & 3 & -7917 & 191  & -8108 \\
     DCSBM & 4 & -7748  & 229  & -7977  \\
     DCSBM & 5 & -7101 & 275  & -7376 \\
     DCSBM & 6 & -6717  & 304  &-7021  \\
     DCSBM & 7 & -6695  & 361 & -7056 \\
     LPCM & 1 & -8662 & 179 & -8841  \\
     LPCM & 2 & -8602  & 184  & -8786  \\
     LPCM & 3 & -8591 & 188 & -8779  \\
     LPCM & 4 & -8493 & 194  & -8687  \\
     LPCM & 5 & -8492  & 205  & -8697  \\
     LPCM & 6 & -8495 & 207 & -8702 \\
     LPCM & 7 & -8502 & 248 & -8750 \\

      \hline\hline
  \end{tabular}
  \end{center}
\end{table}

We note that a zero-dimensional latent space corresponds to a stochastic blockmodel, and a 1-block model corresponds to the latent space model. Based on the results in Table~\ref{table:waic}, a generalized latent space stochastic blockmodel with nine blocks and two-dimensional latent spaces is selected.   

The posterior mean of the block-to-block connection probabilities between the nine blocks calculated from the MCMC samples are provided in Table~\ref{table:block-block}. The off-diagonal entry $(k,l),$ $k \neq l$ represents the probability that a node from block $k$ follows a node from block $l$ when the sender has sender effect zero and receiver has receiver effect zero. The diagonal elements give the range of within block connection probabilities for nodes within the blocks when both nodes have the same latent position.

\begin{table}[htbp]
    \caption{Posterior mean block-to-block connection probabilities.}
    \label{table:block-block}
{%\footnotesize

  \begin{tabular}{lrrrrrr}\hline\hline
  From/ & 1 & 2 & 3 & 4 & 5 & 6 \\
  To & & & & & &  \\ \hline
1&    0.65   & 0.63 & 0.11 & 0.20 & 0.89 & 0.65  \\ 
 2&  0.19 & 0.85 & 0.15 & 0.46 & 0.41 & 0.22  \\ 
 3&      0.11& 0.40 & 0.31 & 0.12 & 0.17 & 0.11  \\
 4&      0.16 & 0.69 & 0.10 & 0.42 & 0.31 & 0.13 \\ 
 5&   0.50 & 0.58 & 0.10 & 0.16 & 0.95 & 0.65 \\ 
 6&     0.30 & 0.41 & 0.05 & 0.06 & 0.65 & 0.61 \\   
  \hline\hline
  \end{tabular}}
\end{table}

\begin{comment}

\begin{table}[htbp]
    \caption{Posterior mean block-to-block connection probabilities. For the within block connections, the range of estimated connection probabilities is shown. Probabilities greater than .5 are highlighted.}
    \label{table:block-block}
{\footnotesize

  \begin{tabular}{lrrrrrrrrr}\hline\hline
  From/ & 1 & 2 & 3 & 4 & 5 & 6& 7 & 8 & 9 \\
  To & & & & & & & & & \\\hline
1&       0.12--1.00 & 0.14 & 0.19 & \cellcolor{light-gray} 0.84 & \cellcolor{light-gray}0.70 & 0.03 & 0.07 & 0.33 & \cellcolor{light-gray}0.68 \\ 
 2&  \cellcolor{light-gray}   0.51 & 0.21--1.00 & 0.30 & 0.06 & 0.06 & 0.09 & 0.14 & 0.30 &\cellcolor{light-gray} 0.86 \\ 
 3&      0.49 & 0.20 & 0.01--1.00 & 0.21 & 0.05 & 0.02 & 0.38 & 0.15 &\cellcolor{light-gray} 0.86 \\
 4&      0.31 & 0.07 & 0.06 & 0.00--1.00 & 0.08 & 0.03 & 0.01 & 0.29 & 0.34 \\ 
 5&   \cellcolor{light-gray}   0.84 & 0.10 & 0.08 & 0.04 & 0.00--1.00 & 0.02 & 0.02 & 0.35 & 0.45 \\ 
 6&      0.38 & 0.41 & 0.20 & 0.16 & 0.09 & 0.00--1.00 & 0.06 & 0.28 & \cellcolor{light-gray}0.65 \\   
 7&   \cellcolor{light-gray}   0.60 & 0.25 &\cellcolor{light-gray} 0.88 & 0.02 & 0.07 & 0.01 & 0.00--0.96 & 0.26 &\cellcolor{light-gray} 0.92 \\ 
 8&    \cellcolor{light-gray}  0.87 & 0.48 & 0.27 & 0.43 & 0.41 & 0.11 & 0.15 & 0.77--1.00 &\cellcolor{light-gray} 0.95 \\ 
 9&    \cellcolor{light-gray}  0.72 &\cellcolor{light-gray} 0.52 &\cellcolor{light-gray} 0.82 & 0.05 & 0.12 & 0.05 & 0.42 & 0.45 & 0.54--1.00 \\ \hline\hline
  \end{tabular}}
\end{table}

\end{comment}

It can be seen that the block-to-block connection probabilities exhibit considerable asymmetry.  For example, the probability that a member of block 1 connects with block 2 is $\tau_{12}=0.63$, whereas the probability that a member of block 2 connects with block 1 is $\tau_{21}=0.19$. 

Furthermore, some between-block connection probabilities give evidence of disassortative mixing. For example, some politicians within block 4 have probability $0.42$ of connecting with each other, whereas the probability of a politician from block 4 connecting with a politician from block 2 is $0.69$. Likewise, some politicians in block 6 have probability $0.61$ of connecting with each other, whereas the probability of them connecting with a politician from block 5 is $0.65$.

\begin{figure}[htbp]
\begin{center}
\includegraphics[width=0.9\textwidth]{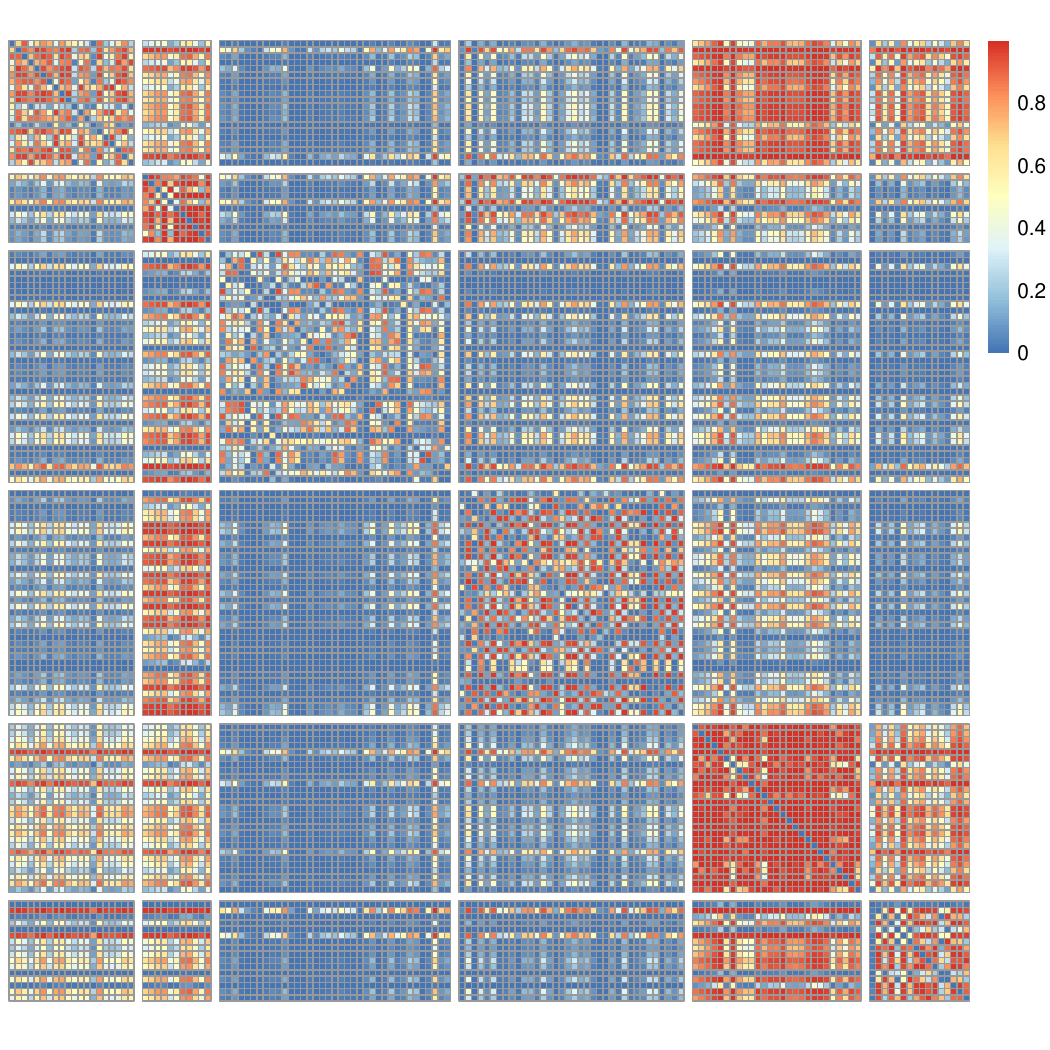}
\end{center}
\caption{A heat map representation of the edge probability matrix for all of the politicians.}\label{fi:Pmatrix}
\end{figure}

A heat map of the edge probabilities is given in Figure~\ref{fi:Pmatrix} and this shows the joint effect of block membership, latent positions, sender and receiver effects for each politician. The heat map shows a strong block structure in the edge probabilities where members within blocks have similar, but non identical, probability of following other members of other blocks. 

The posterior block allocation probabilities are presented in Table~\ref{table:block_allocation}, where we can see that blocks 3 and 4 have higher membership probability than the other blocks, whereas block 2 has the lowest membership probability.

\begin{table}[htbp]

    \caption{Posterior mean of the block allocation probabilities.}
    \label{table:block_allocation}

\begin{center}  \begin{tabular}{rrrrrrrrr}\hline\hline
  \multicolumn{6}{c}{Block}\\
  1 & 2 & 3 & 4 & 5 & 6\\\hline
    0.14 & 0.10 & 0.23 & 0.22 & 0.18 & 0.13  \\ \hline\hline
  \end{tabular}\end{center}
\end{table}

\begin{comment}

\begin{table}[htbp]

    \caption{Posterior mean of the block allocation probabilities.}
    \label{table:block_allocation}

\begin{center}  \begin{tabular}{rrrrrrrrr}\hline\hline
  \multicolumn{9}{c}{Block}\\
  1 & 2 & 3 & 4 & 5 & 6 & 7 & 8 & 9\\\hline
     0.07 & 0.15 & 0.19 & 0.15 & 0.09 & 0.08 & 0.14 & 0.06 & 0.08  \\ \hline\hline
  \end{tabular}\end{center}
\end{table}

\begin{table}[ht]
    \caption{Average receiver and sender effects for members in each political party.
    }
    \label{table:dail_degree}
\smallskip
\centering
{ \begin{tabular}{lrr}
  \hline\hline
Party & Average receiver effects & Average sender effects \\ 
  \hline
Fianna F\'{a}il &  -0.42 &  -1.20 \\ 
  Fine Gael &  0.12 & 0.14 \\ 
  Independent &  0.65 & 0.03 \\ 
  Labour &  0.27 &  0.76 \\ 
  People Before Profit Alliance & -4.67 &  0.51 \\ 
  Renua Ireland & 1.42 & 0.52 \\ 
  Sinn F\'{e}in & -0.56 &  -1.26 \\ 
  Social Democrats & 0.39 &  2.13 \\ 
  Socialist Party &  -1.80 & -1.12 \\ 
  United Left &   -3.08 & -0.22 \\ 
   \hline\hline
\end{tabular}}
\end{table}

\end{comment}

Table~\ref{table:political_party} presents the relationships between block memberships and political party memberships.

\begin{table}[htbp]

    \caption{Political party and maximum {\em a posteriori} block membership.}
    \label{table:political_party}

\begin{center}
{\small \begin{tabular}{rrrrrrr}
  \hline\hline
 & 1 & 2 & 3 & 4 & 5 & 6  \\ \hline
Fianna F\'{a}il &   - &  2 &   - &   15 &   - &   -  \\ 
  Fine Gael &   18 &   - &  2 &   1 &   27 &  13 \\ 
  Independent &   1 &   1 &   12 &   1 &   - &   2\\ 
  Labour &   - &  8 &   2 &   19 &   - &   -  \\ 
  People Before Profit Alliance &   - &   - &   1 &   - &   - &   - \\ 
  Renua Ireland &   1 &   - &   - &   - &   - &   1  \\ 
  Sinn F\'{e}in &   - &   - &   12 &  - &   - &   - \\ 
  Social Democrats &   - &   - &   3 &   - &   - &   -  \\
  Socialist Party &   - &   - &   3 &   - &   - &   -\\ 
  United Left &   - &   - &   2 &   - &   - &   -  \\ 
   \hline\hline
\end{tabular}}
\end{center}
\end{table}

\begin{comment}
\begin{table}[htbp]

    \caption{Political party and maximum {\em a posteriori} block membership.}
    \label{table:political_party}

\begin{center}
{\small \begin{tabular}{rrrrrrrrrr}
  \hline\hline
 & 1 & 2 & 3 & 4 & 5 & 6 & 7 & 8 & 9 \\ \hline
Fianna F\'{a}il &   5 &   - &   - &   2 &   9 &   - &   - &   1 &   - \\ 
  Fine Gael &   - &   - &  30 &   1 &   - &   2 &  21 &   - &   7 \\ 
  Independent &   2 &   1 &   2 &   5 &   2 &   2 &   2 &   1 &   - \\ 
  Labour &   - &  24 &   - &   - &   - &   3 &   - &   - &   2 \\ 
  People Before Profit Alliance &   - &   - &   - &   1 &   - &   - &   - &   - &   - \\ 
  Renua Ireland &   - &   - &   2 &   - &   - &   - &   - &   - &   - \\ 
  Sinn F\'{e}in &   - &   - &   - &  12 &   - &   - &   - &   - &   - \\ 
  Social Democrats &   - &   - &   - &   - &   - &   1 &   - &   2 &   - \\
  Socialist Party &   - &   - &   - &   3 &   - &   - &   - &   - &   - \\ 
  United Left &   - &   - &   - &   1 &   - &   1 &   - &   - &   - \\ 
   \hline\hline
\end{tabular}}
\end{center}
\end{table}
\end{comment}

The two parties in government (Fine Gael and Labour) largely belong to blocks 1, 5, 6 and 2, 4 respectively whereas all of the members of the opposition party, Sinn F\'{e}in are assigned to block 3, along with most of the independent and small parties. Interestingly, Fianna F\'{a}il are mainly assigned to block 4 which contains a large number of Labour politicians.

The three blocks dominated by Fine Gael politicians have average block-to-block interaction probabilities greater than $0.30$. There is asymmetry in the block-to-lock interaction probabilities, particularly $\tau_{15}=0.89$ whereas $\tau_{51}=0.50$ and $\tau_{16}=0.65$ whereas $\tau_{61}=0.30$. This asymmetry in following relationships can explain why the Fine Gael politicians are divided across multiple blocks. 

By examining the maximum {\em a posteriori} block membership of each party member, we developed text descriptors for each block; these are given in Table~\ref{table:descriptors}.
 
\begin{table}[ht]
\caption{Text descriptors of the blocks in the chosen model. }
\label{table:descriptors}
\begin{center}
\begin{tabular}{lr}\hline\hline
Block & Descriptor \\\hline
1 &  Fine Gael\\
2 &  Labour with Fianna F\'{a}il \\
3 &  Sinn F\'{e}in, Independent and small parties\\
4 &  Labour and Fianna F\'{a}il\\
5 &  Fine Gael\\
6 & Fine Gael\\ \hline\hline
\end{tabular}
\end{center}
\end{table}

We can see that the block membership is largely dictated by party membership.

\begin{figure}[htbp]
\begin{tabular}{cc}
\includegraphics[width=0.55\textwidth]{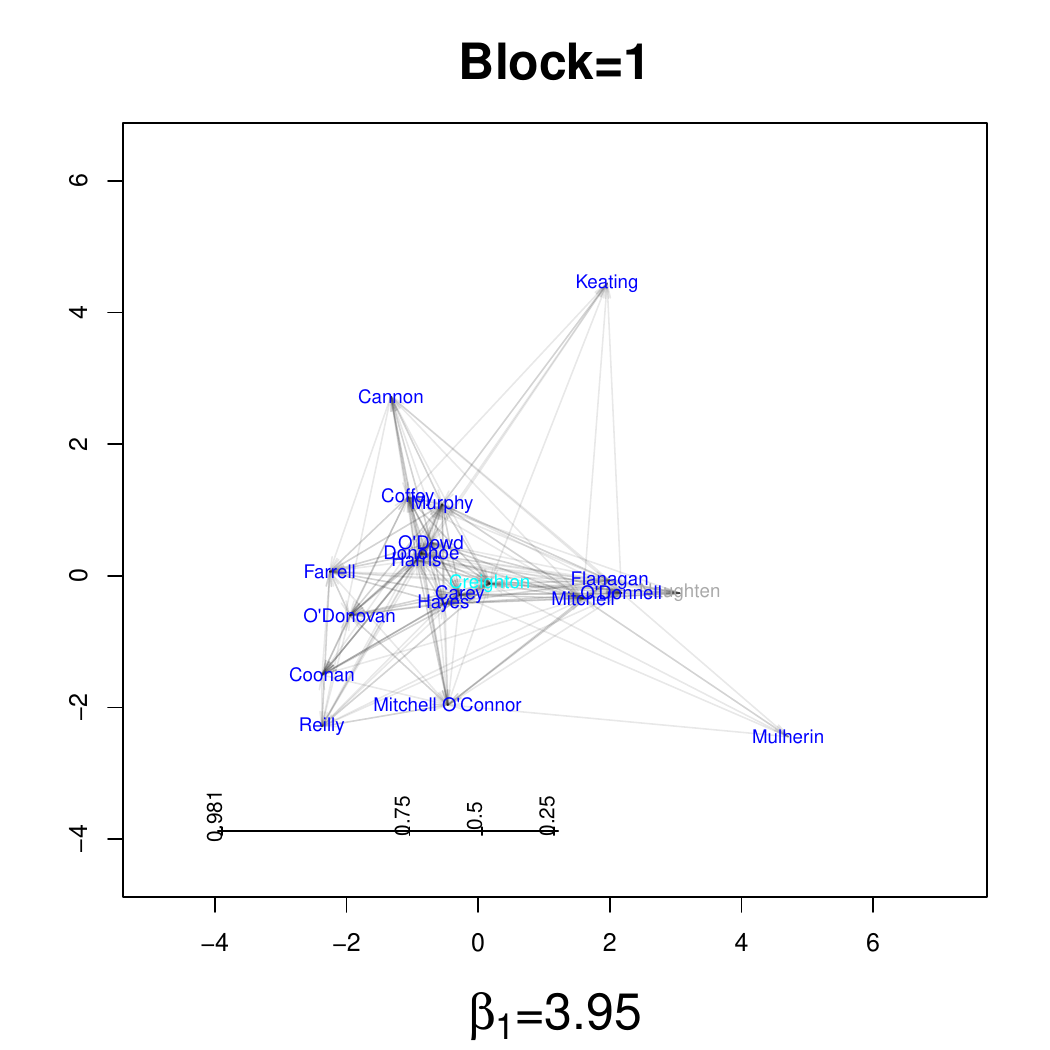}
 & \includegraphics[width=0.55\textwidth]{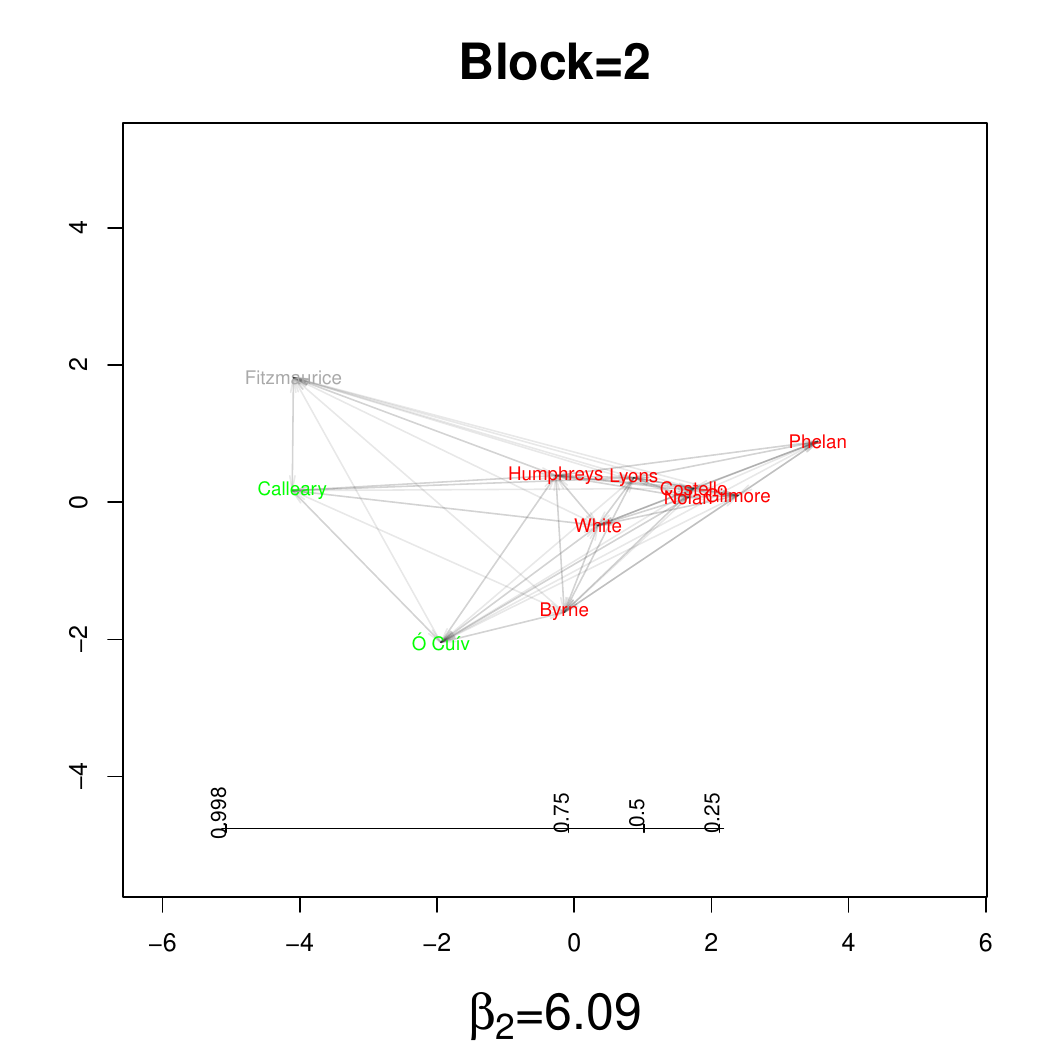}
\\

\includegraphics[width=0.55\textwidth]{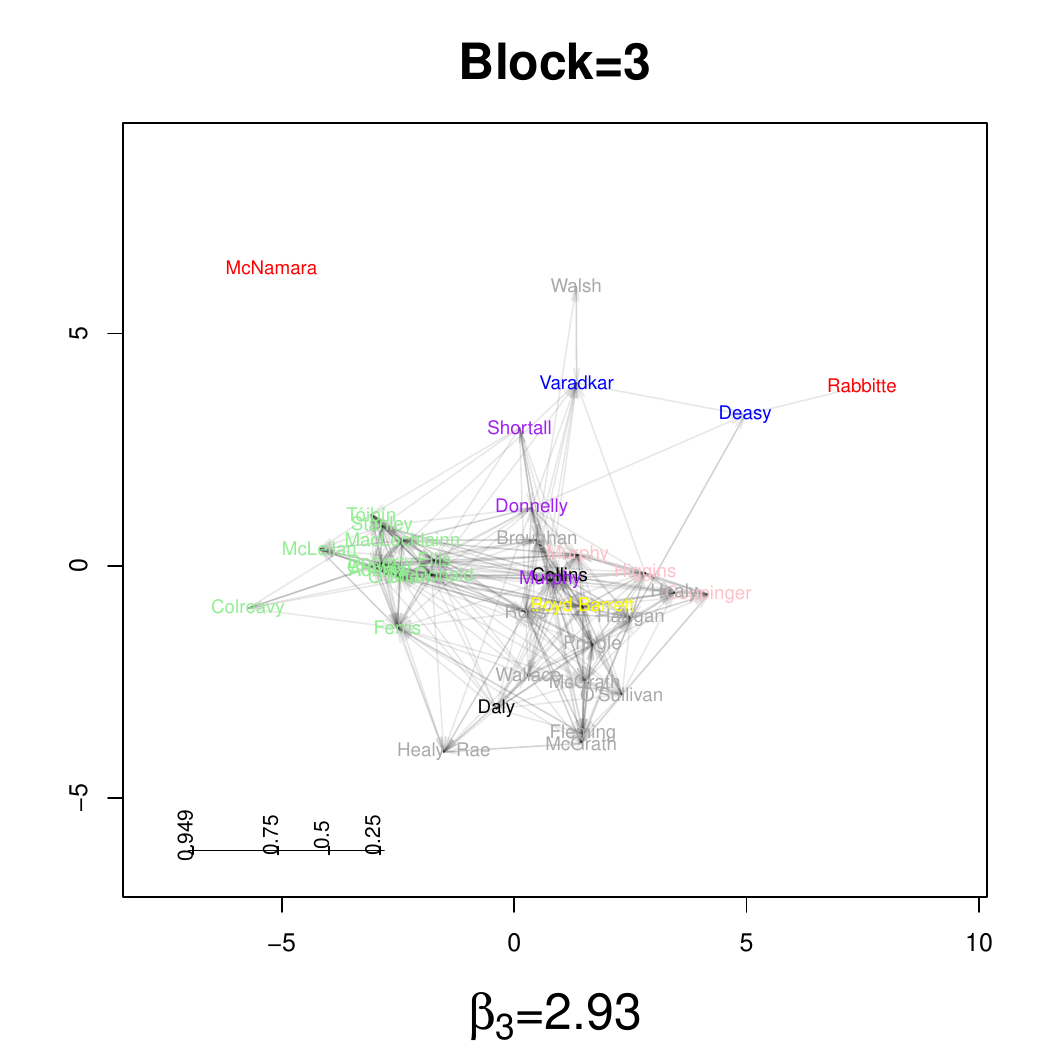}
&
\includegraphics[width=0.55\textwidth]{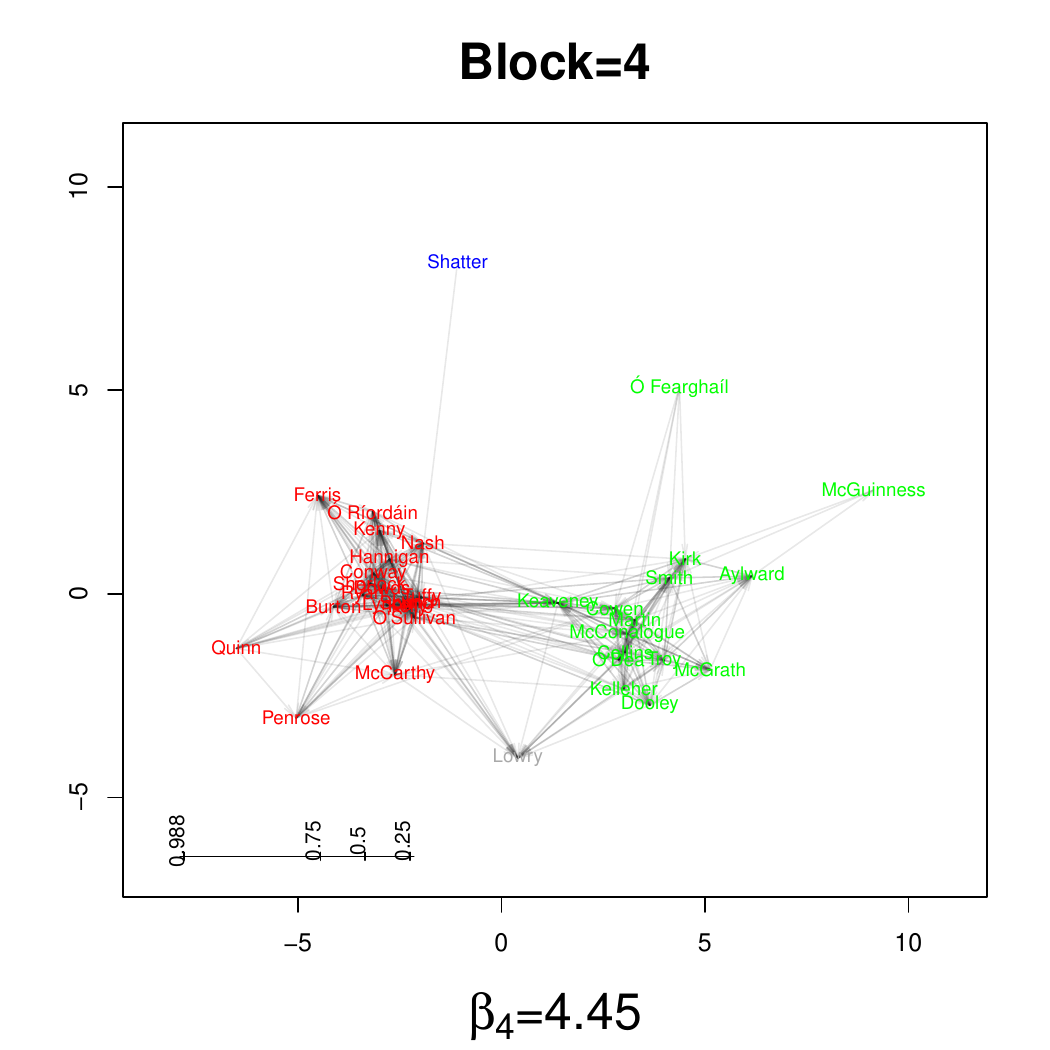}
\\

\includegraphics[width=0.55\textwidth]{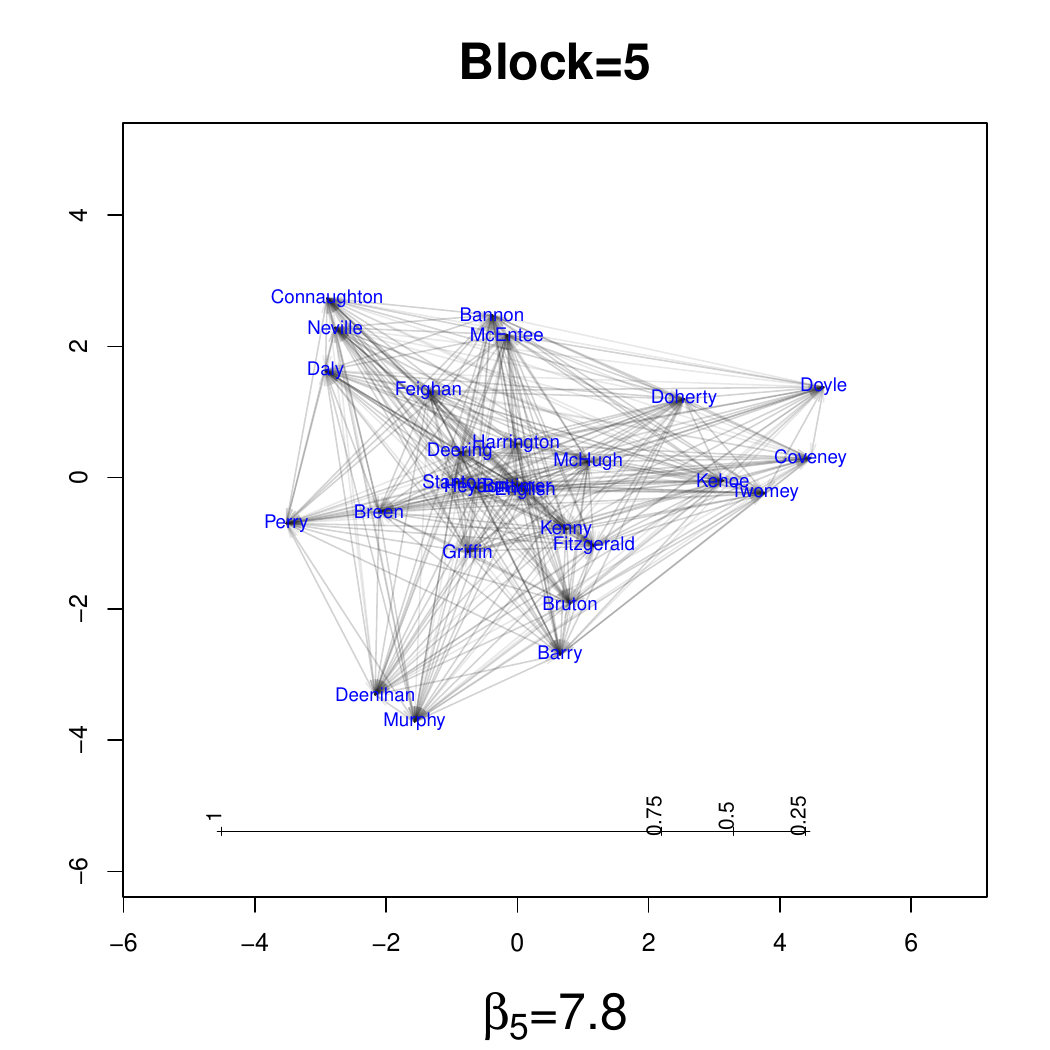}
&
\includegraphics[width=0.55\textwidth]{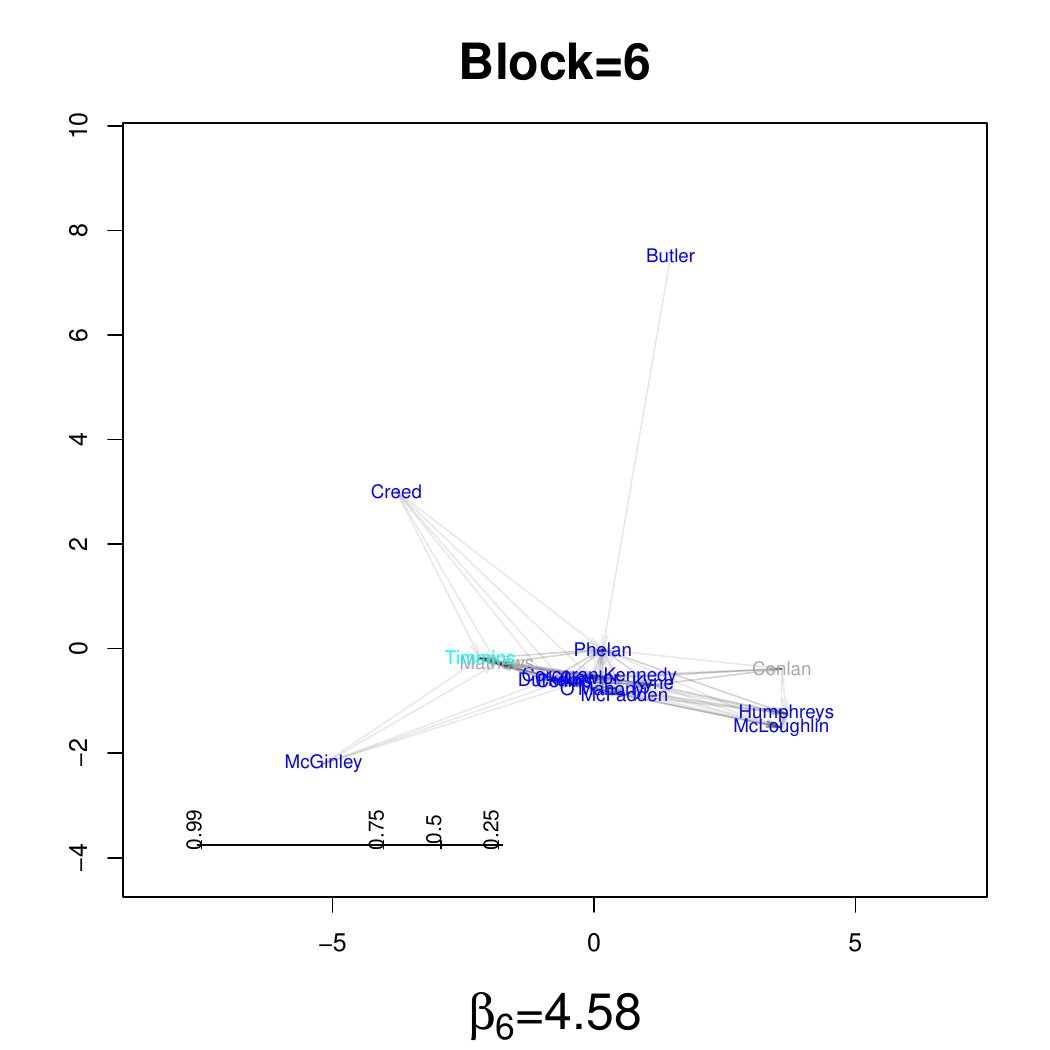}
\\

\end{tabular}
\caption{Posterior mean latent positions within each block. The politicians are colored by political party. A scale has been added to each plot to show how the link probabilities depend on distance; the left of the scale is the highest probability and the probability decreases as the distance from the left increases on the scale.}
\label{fig:positions}
\end{figure}

The posterior mean latent positions for each node within its maximum {\em a posteriori} block membership is shown in Figure~\ref{fig:positions}. The locations of nodes within the block allow us to see additional structure among politicians, beyond what is possible with a stochastic blockmodel.

\begin{comment}
\begin{figure}[htbp]
\begin{tabular}{ccc}
\includegraphics[width=0.3\textwidth]{Dail/Zlocnew-scale1.pdf}
 & \includegraphics[width=0.3\textwidth]{Dail/Zlocnew-scale2.pdf}
&
\includegraphics[width=0.3\textwidth]{Dail/Zlocnew-scale3.pdf}
\\

\includegraphics[width=0.3\textwidth]{Dail/Zlocnew-scale4.pdf}
&
\includegraphics[width=0.3\textwidth]{Dail/Zlocnew-scale5.pdf}
&
\includegraphics[width=0.3\textwidth]{Dail/Zlocnew-scale6.pdf}
\\

\includegraphics[width=0.3\textwidth]{Dail/Zlocnew-scale7.pdf}
&
\includegraphics[width=0.3\textwidth]{Dail/Zlocnew-scale8.pdf}
&
\includegraphics[width=0.3\textwidth]{Dail/Zlocnew-scale9.pdf}\\

\end{tabular}
\caption{Posterior mean latent positions within each block. The politicians are colored by political party. A scale has been added to each plot to show how the link probabilities depend on distance; the left of the scale is the highest probability and the probability decreases as the distance from the left increases on the scale.}
\label{fig:positions}
\end{figure}
\end{comment}

One Renua member and an independent politician are located within block 1, which is dominated by Fine Gael. These members were former members of the party and at the time of the election were members of Fine Gael. These politicians either left or were expelled from the party during the sitting of the 31st D\'{a}il; thus these independents and Renua members closely correspond to the apostate independent category of \cite{weeks09} because they are following the party that they were affiliated with in the past. Similarly, in block 6, Timmins, Matthews and Conlon are all former member of Fine Gael who were either expelled or left the party. 

Block 5 consists entirely of Fine Gael members and many of the most senior party members are in this block. The probabilities of blocks 1 and 6 connecting with this block are high, so members of those blocks have a tendency to follow members in this block. 

Membership of block 3 consists mainly of Sinn F\'{e}in members, Independent politicians and small parties. The latent space layout shows that within the block there is a tendency of similar politicians to be located close together. Surprisingly two Fine Gael politicians have been assigned to this block, including Varadkar who subsequently (June 2nd, 2017) became the leader of Fine Gael. 

Finally, blocks 2 and 4 consist mainly of Fianna F\'{a}il and Labour politicians. Block 2 is more closely connected and members have a higher probability of following members of block 4 than vice versa. 

\begin{figure}[htbp]
\begin{center}
\includegraphics[width=0.9\textwidth]{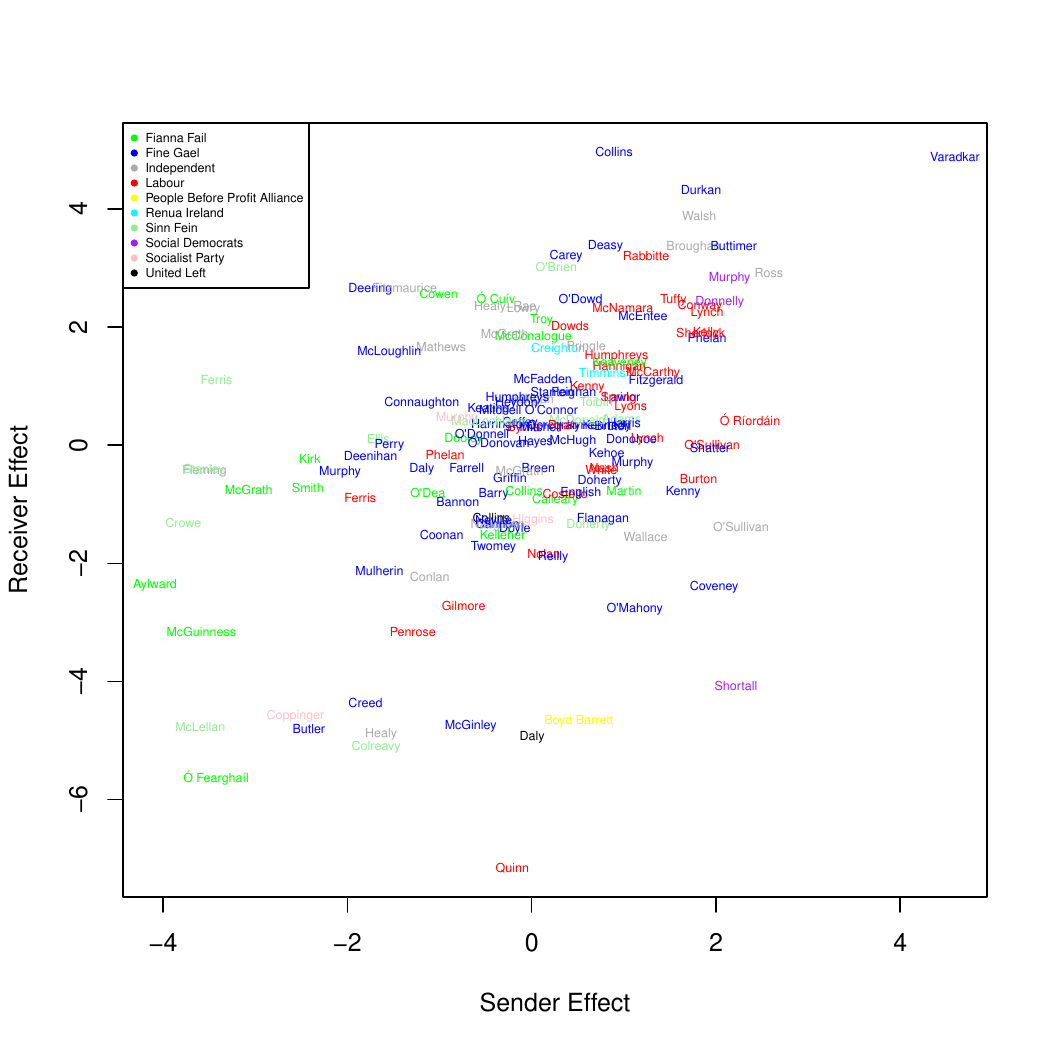}
\end{center}
\caption{A scatter plot of the sender and receiver effects for all of the politicians. The points are coloured by the political party of the politician.}\label{fi:sendreceive}
\end{figure}

The estimated sender and receiver effects of the politicians are shown in Figure~\ref{fi:sendreceive}.

Leo Varadkar has a very high sender and receiver effect. Thus, he has a higher probability of following and being followed by politicians when compared to other members of the block to which he is assigned (Block 3). Thus, even though he has been assigned to a different block than many members of his political party, he still has a high probability of connecting them them. 

As a next step in the analysis, we considered each vote in the D\'{a}il and grouped all pairs of politicians into three groups on the basis of their voting agreement: agree, disagree or at least one was absent. Kernel density estimates of the edge probabilities were formed for each group and each vote. The resulting plot for one vote is shown in Figure~\ref{fig:Voting}; the equivalent plot for the remaining 22 votes has similar structure (see Appendix~\ref{se:KDE}).

\begin{figure}[htb]
\begin{center}
\includegraphics[width=0.6\textwidth]{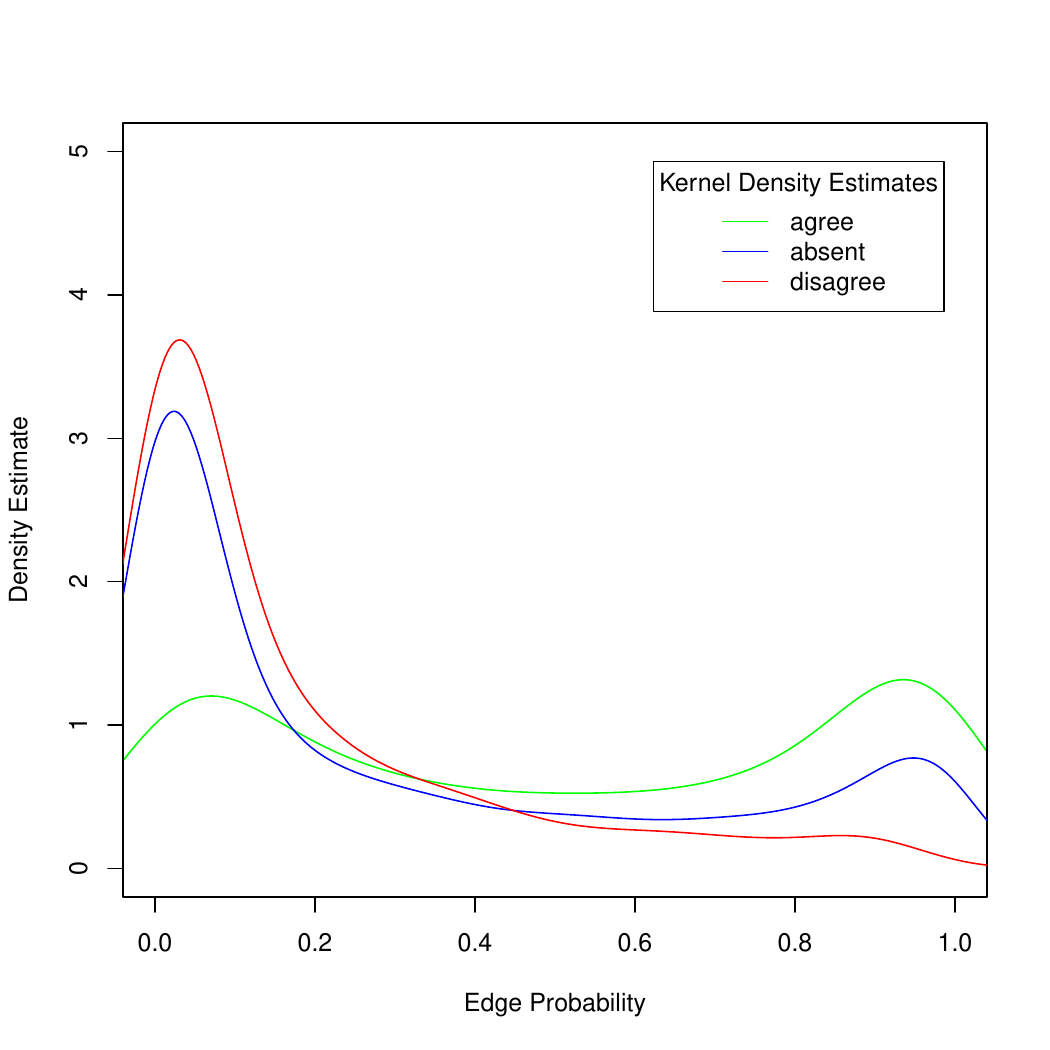}
\end{center}
\caption{A comparison of the edge probabilities from the model for politicians grouped by voting agreement/disagreement.}\label{fig:Voting}
\end{figure}

We can see from the plot that the politicians who agree in the D\'{a}il votes tend to have higher edge probabilities than those who disagree, with the absent group having intermediate probabilities. Thus, the model edge probabilities have some predictive importance in terms of voting in the D\'{a}il. This result can be partially explained by the effect of political party on Twitter following and on the effect of political party on voting in the D\'{a}il \citep{farrell15}. However, the result helps us understand the role that independent politicians play in the D\'{a}il.

\begin{figure}[htbp]
\begin{center}
\includegraphics[width=0.8\textwidth]{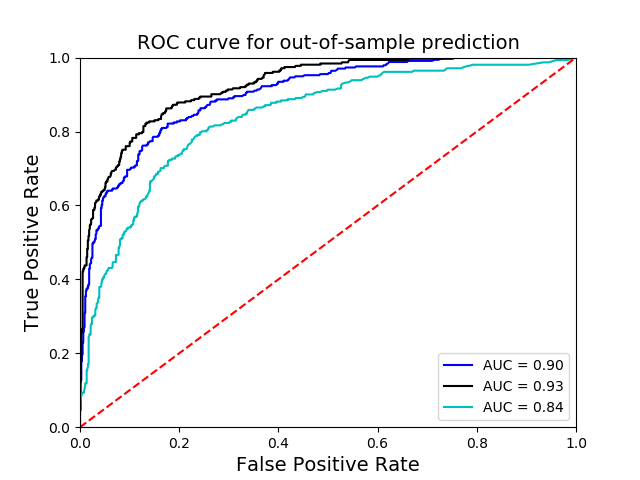}
\end{center}
\caption{ROC curve for out-of-sample predictions using a 6-block deegree-corrected stochastic blockmodel (blue), a 4-cluster latent position cluster model (green) and a 6-block GLSSBM (black).} \label{fig:ROC}
\end{figure}

Finally, we completed a 10-fold cross-validation study to assess the model performance for edge prediction. The results are shown in a box plot in Figure~\ref{fig:ROC}, where it can be seen that the edge probabilities from the GLSSBM model are better for edge prediction than the stochastic blockmodel, but the latent space model is very competitive with the GLSSBM model.\begin{comment}
\bay{(How many clusters were specified for the LPCM? Also 9?)} \bay{(We might consider a precision recall curve also.  I recall in another project that should better results because it weighted edges and non-edges more equally.)}
\end{comment}

\section{Discussion}
\label{se:discussion}

A new model has been proposed in this paper, namely, the generalized latent space stochastic blockmodel. The model generalizes both the latent space model  as well as the stochastic blockmodel and simultaneously captures transitivity, clustering and disassortative mixing of networks. A fully Bayesian approach using MCMC sampling was used to estimate the class memberships, latent positions, and model parameters. 

The model has enabled discovering interesting structure in the Irish politician Twitter following network. The resulting estimated model parameters, estimated block memberships and estimated latent positions reveal the relationships between political parties, within political parties and the relationship between independent politicians and small party politicians and the main parties of the D\'{a}il. Most interestingly, many apostate independents remained a member of the block dominated by the party for which they are former members. 

The generalized latent space stochastic blockmodel can be extended in a number of ways. It could be extended to model integer-valued networks by assuming the number of edges between each pair follows a Poisson or negative binomial distribution. Furthermore, covariates could be incorporated into the model through the use of node-specific covariates and edge covariates. 

Additionally, although Bayesian estimation with MCMC sampling is highly flexible, it is computationally prohibitive and unsuitable for larger networks. If the main interest is in the posterior mode, variational methods \citep[eg.][]{smidl06,bishop06} may be more appropriate; this approach to inference is well established some other network models \citep[eg,][]{airoldi06,SalterTownshend2013}.

\paragraph{Acknowledgements}

The authors would like to thank the Editor, Associate Editor and Reviewers for their useful comments that improved this work. We would also like to thank Prof. David Farrell for his advice on relevant research results in Irish politics. 

This work was partially supported by the National Science Foundation under Grant Number SES-1461495 to Fosdick and Grant Numbers SES-1559778 and DMS-1737673 to McCormick.  McCormick is also supported by grant number K01 HD078452 from the National Institute of Child Health and Human Development (NICHD).  This material is based upon work supported by, or in part by, the U. S. Army Research
Laboratory and the U. S. Army Research Office under contract/grant number W911NF-12-1-0379.  Murphy and Ng are supported by the Science Foundation Ireland funded Insight Research Centre (SFI/12/RC/2289\_P2).  The authors would also like to thank the Isaac Newton Institute Program on Theoretical Foundations for Statistical Network Analysis workshop on Bayesian Models for Networks, supported by EPSRC grant number EP/K032208/1. 

\bibliographystyle{ims}
\bibliography{refs}

%\addbibresource{refs.bib}

%\printbibliography

\newpage
\appendix
\section{Kernel Density Estimates for Edge Probabilities}\label{se:KDE}

The kernel density estimates of the edge probabilities for each of the votes from January 1st, 2016 to February 6th, 2016. 

\begin{tabular}{ccc}
Vote 1 & Vote 2 & Vote 3\\
\includegraphics[width=0.3\textwidth]{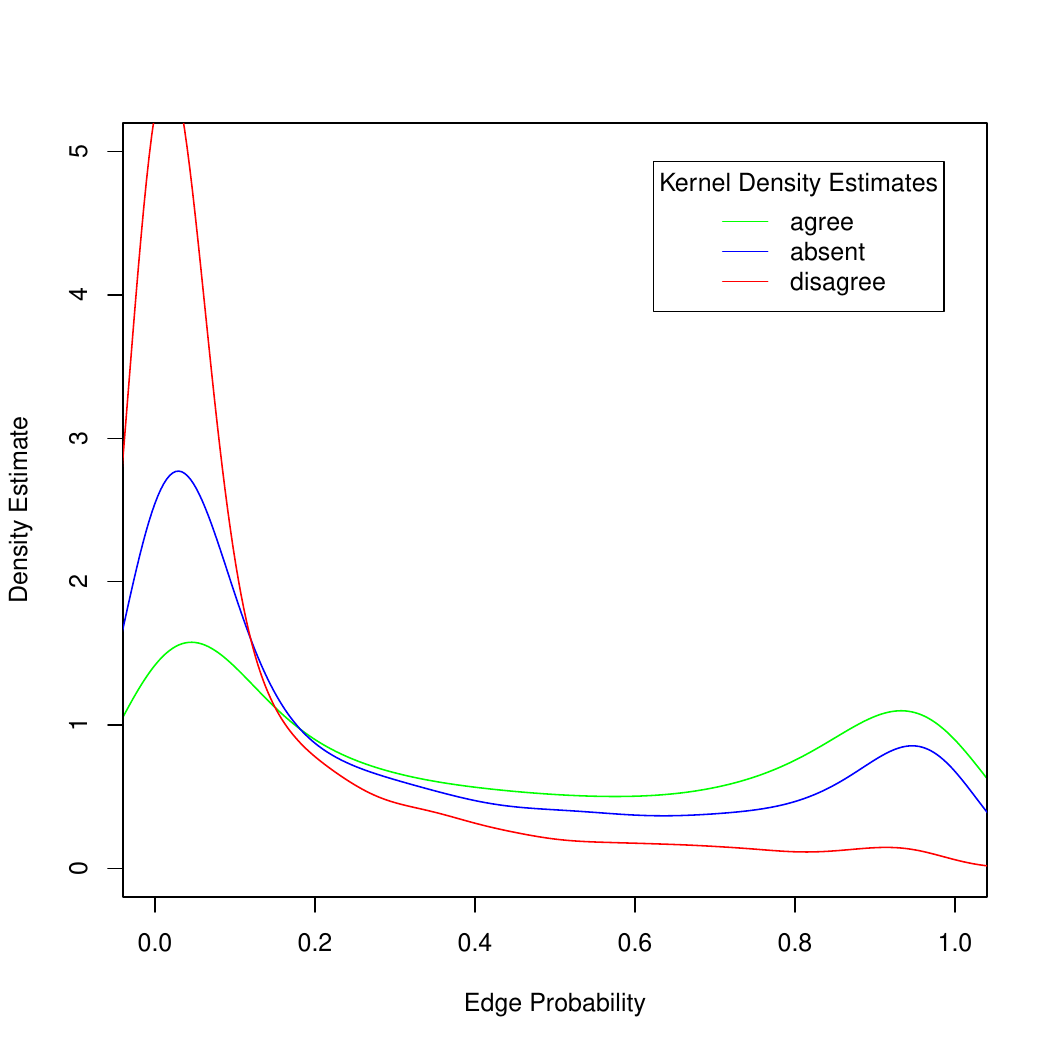} & \includegraphics[width=0.3\textwidth]{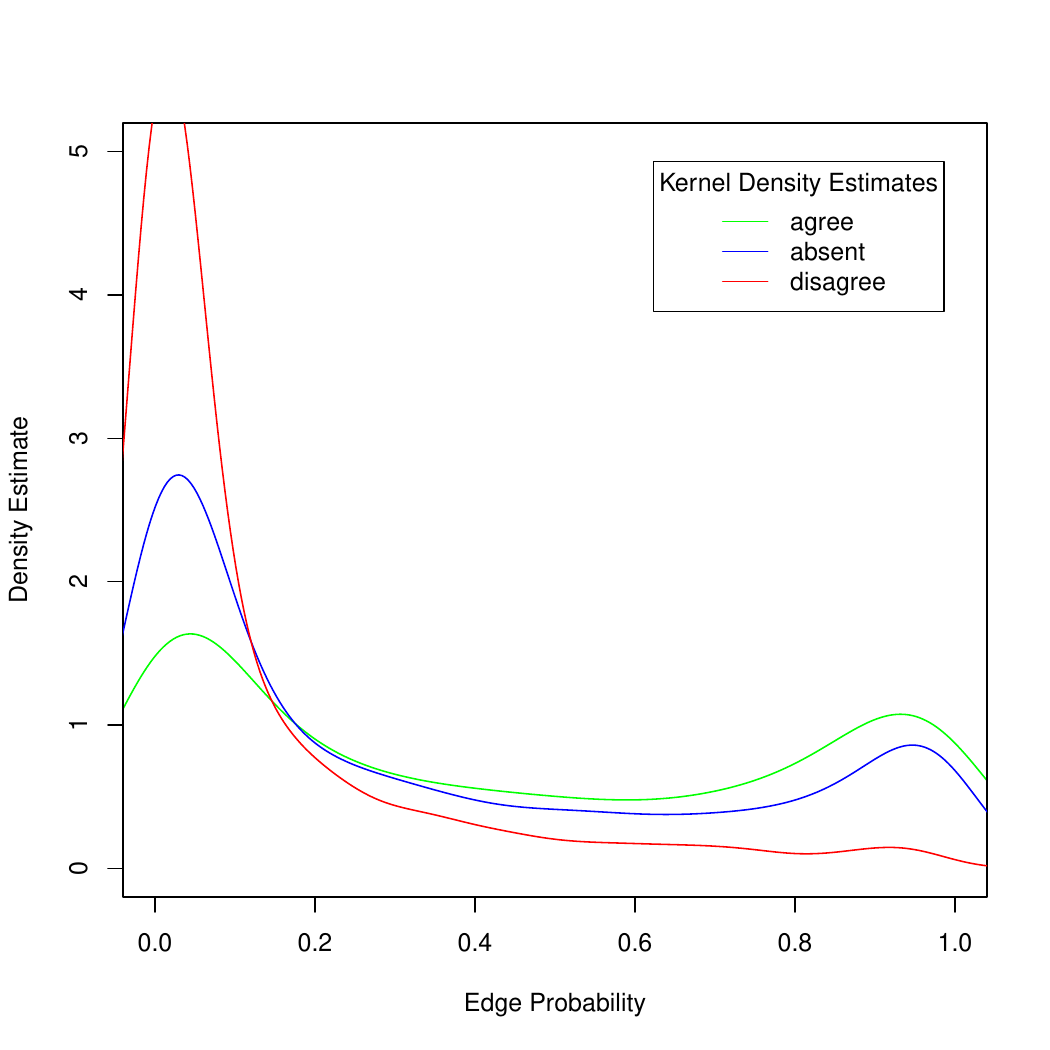} & 
\includegraphics[width=0.3\textwidth]{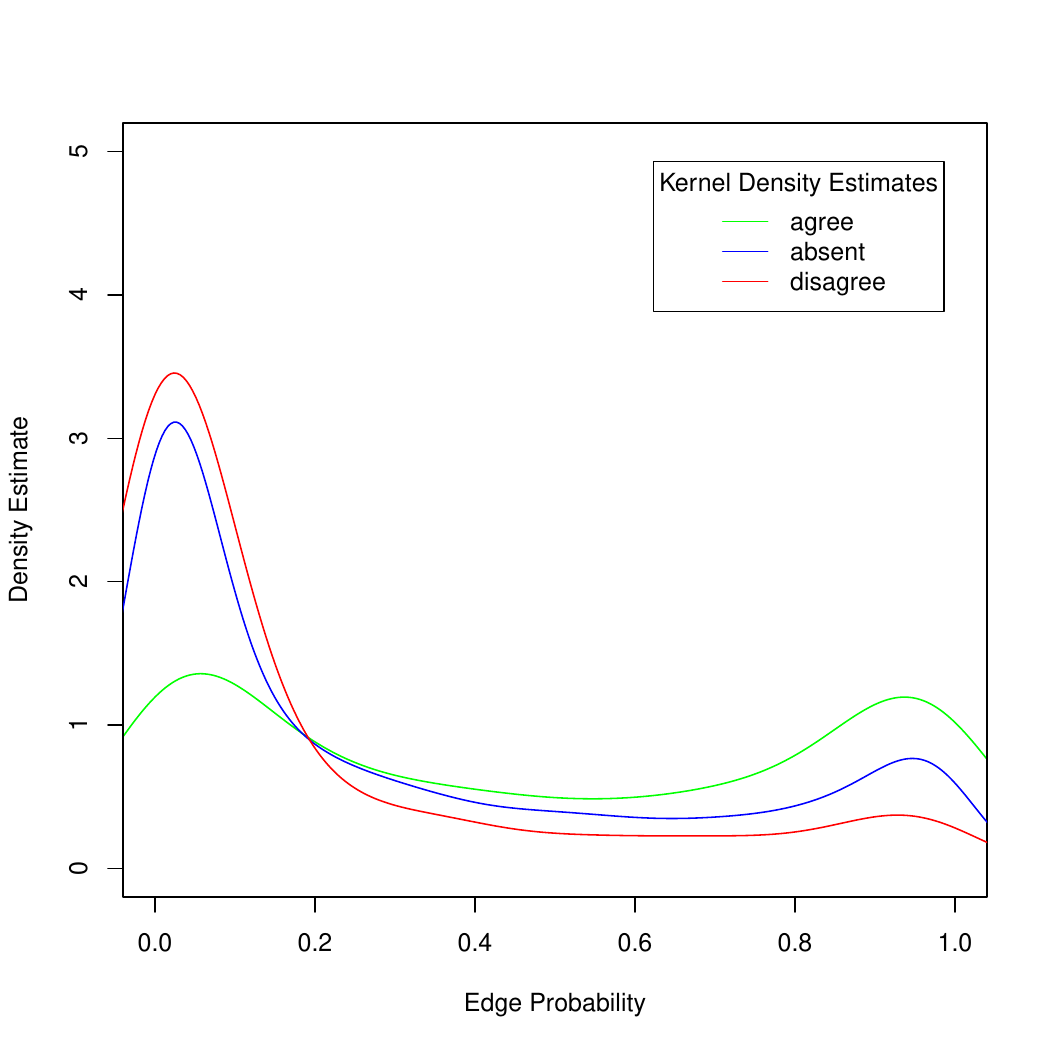} \\
Vote 4 & Vote 5 & Vote 6\\
\includegraphics[width=0.3\textwidth]{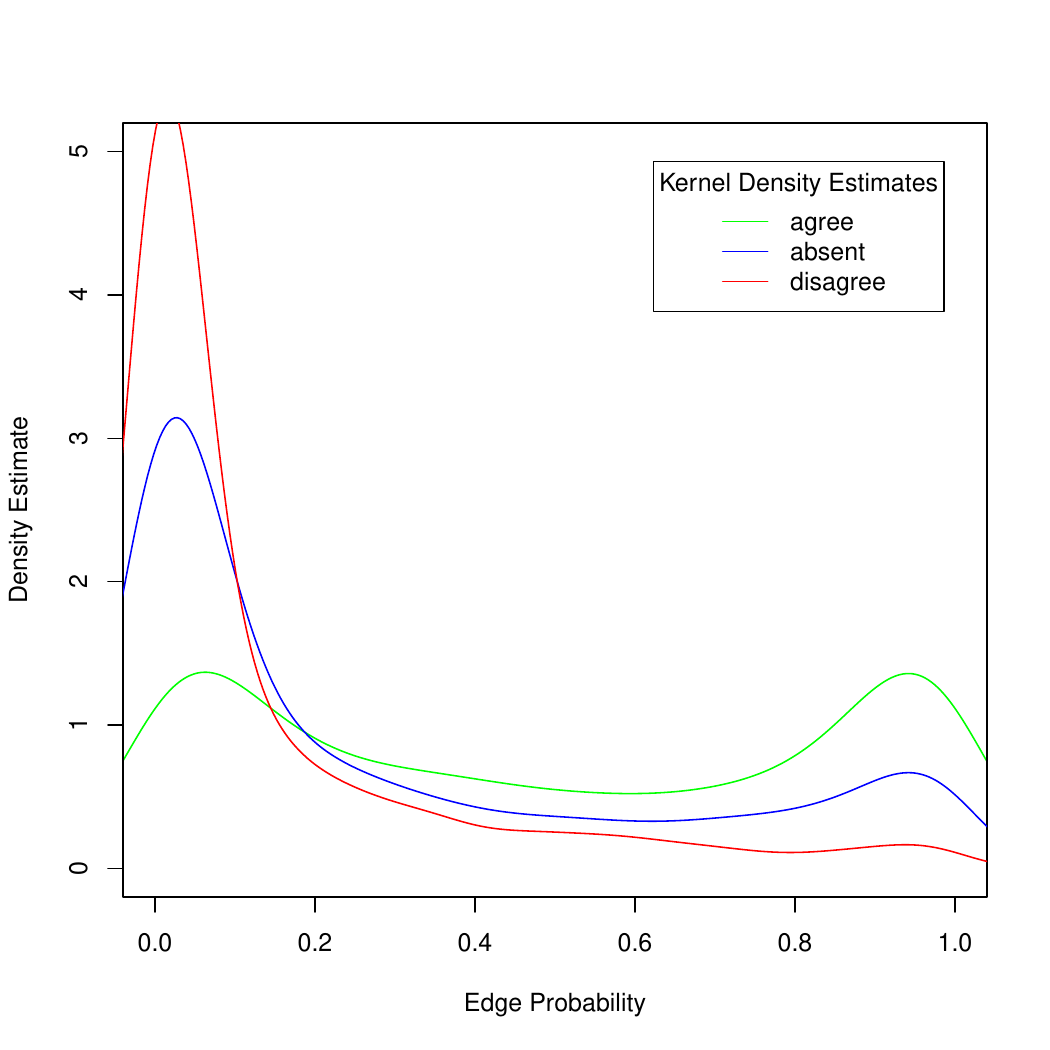} & \includegraphics[width=0.3\textwidth]{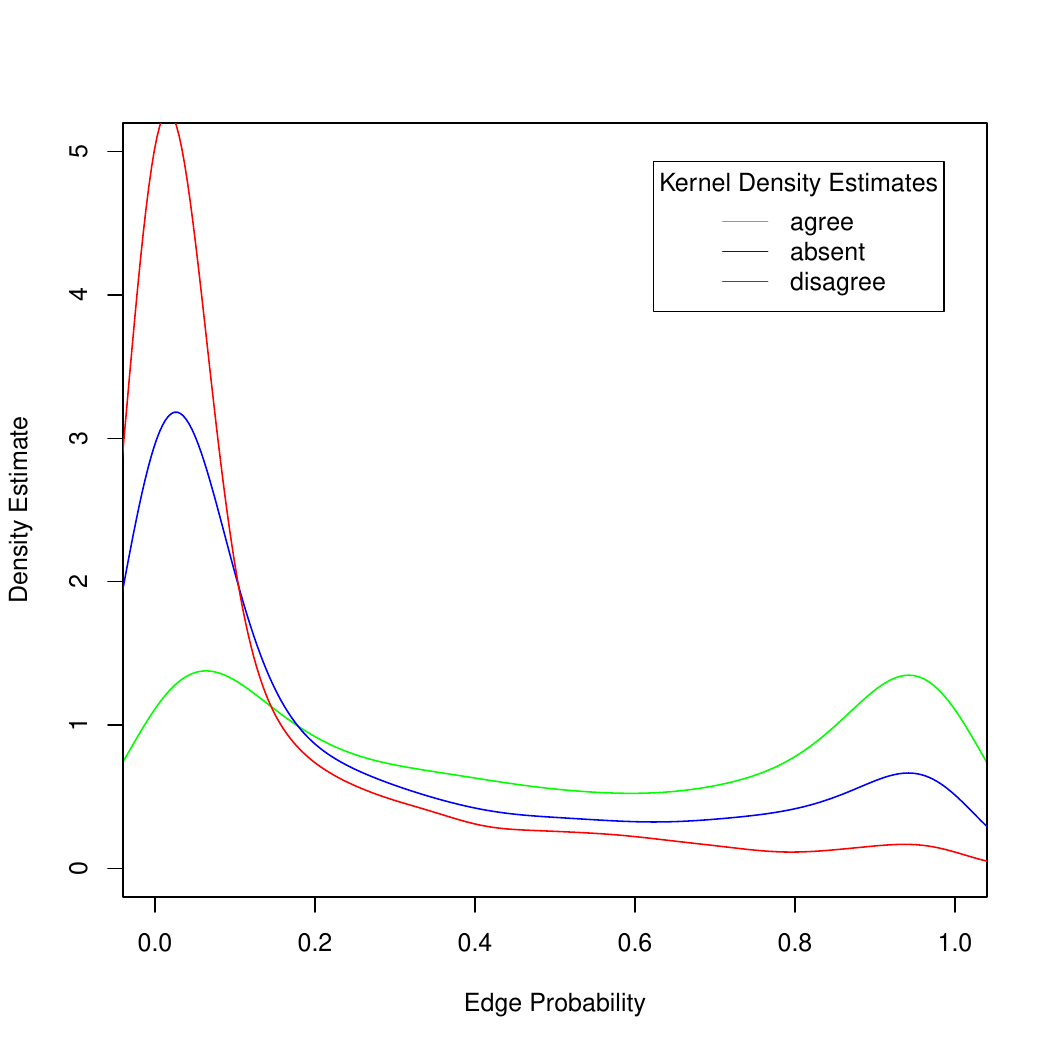} & 
\includegraphics[width=0.3\textwidth]{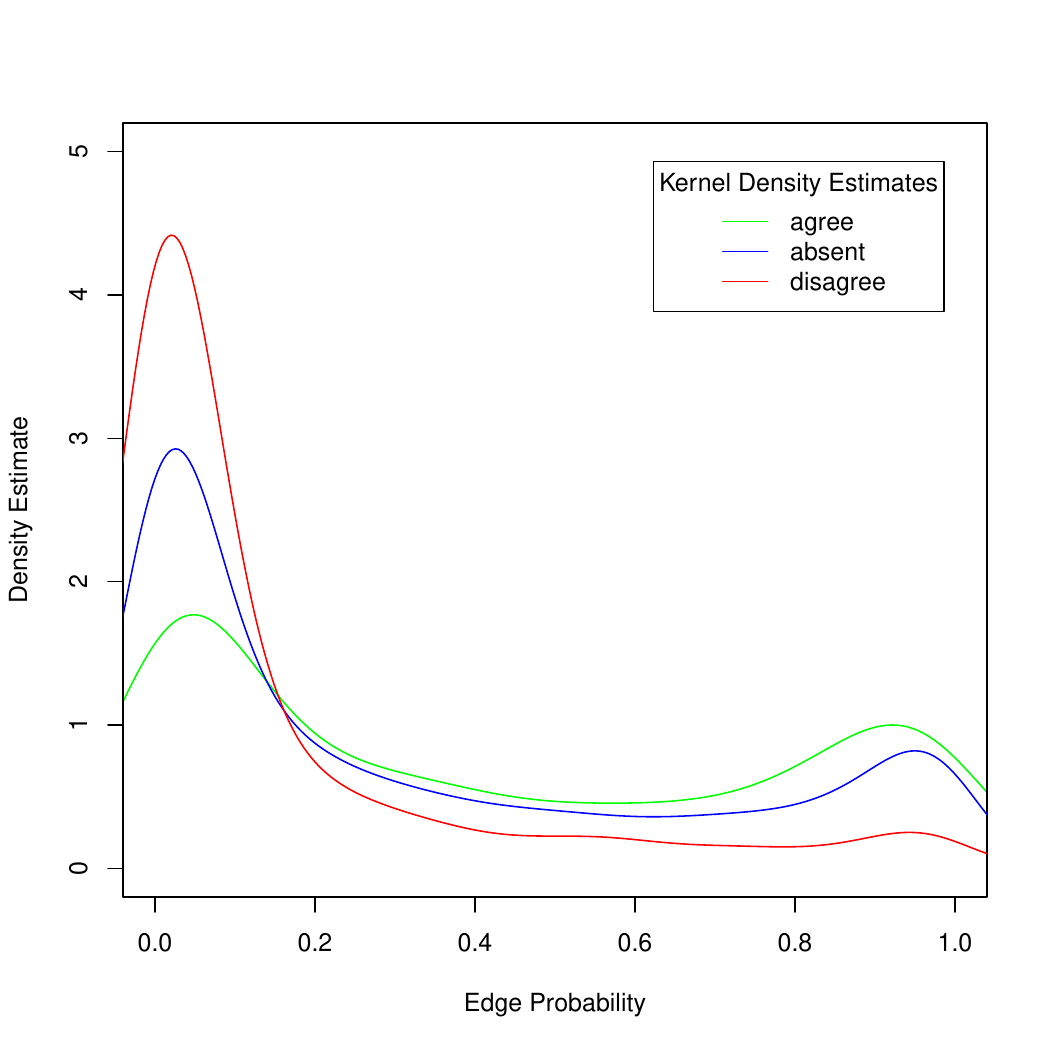} \\
Vote 7 & Vote 8 & Vote 9\\
\includegraphics[width=0.3\textwidth]{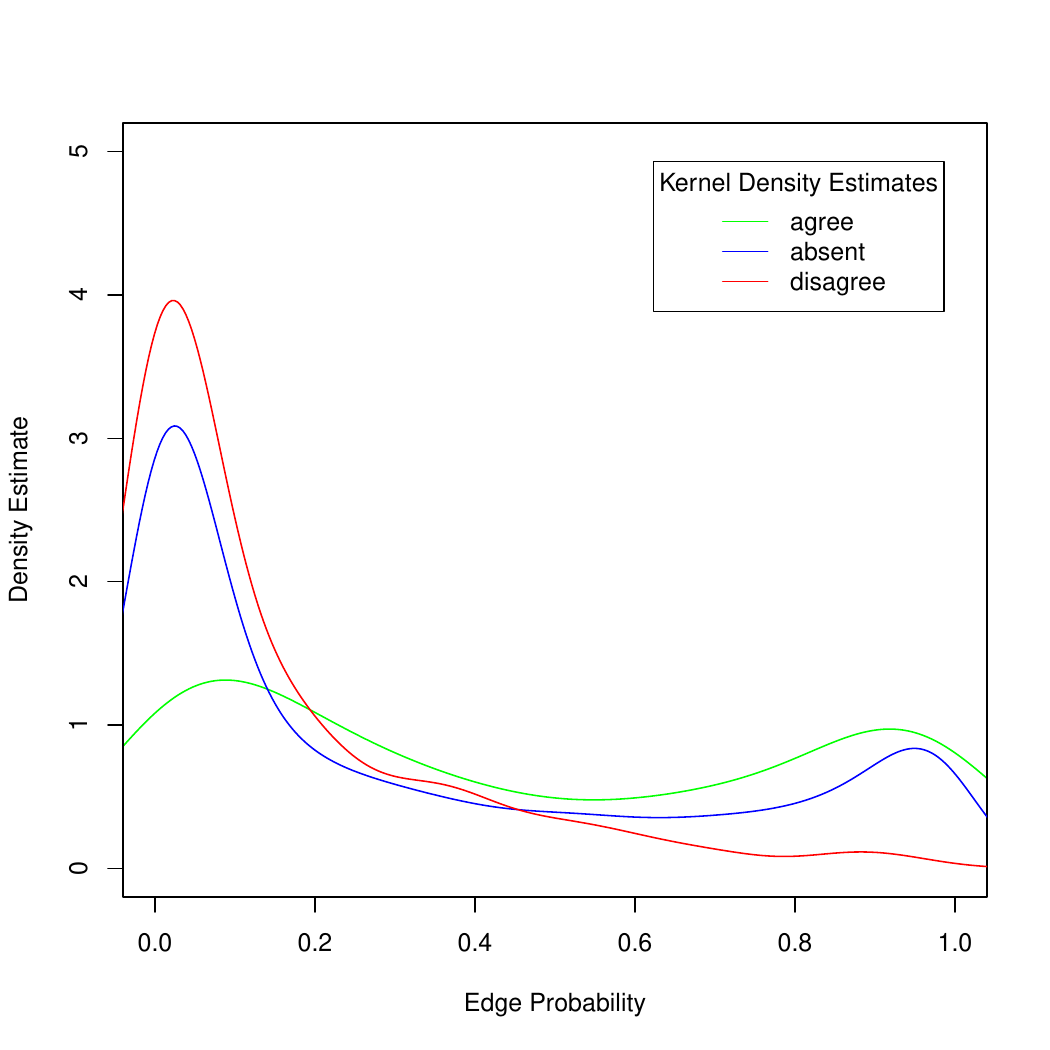} & \includegraphics[width=0.3\textwidth]{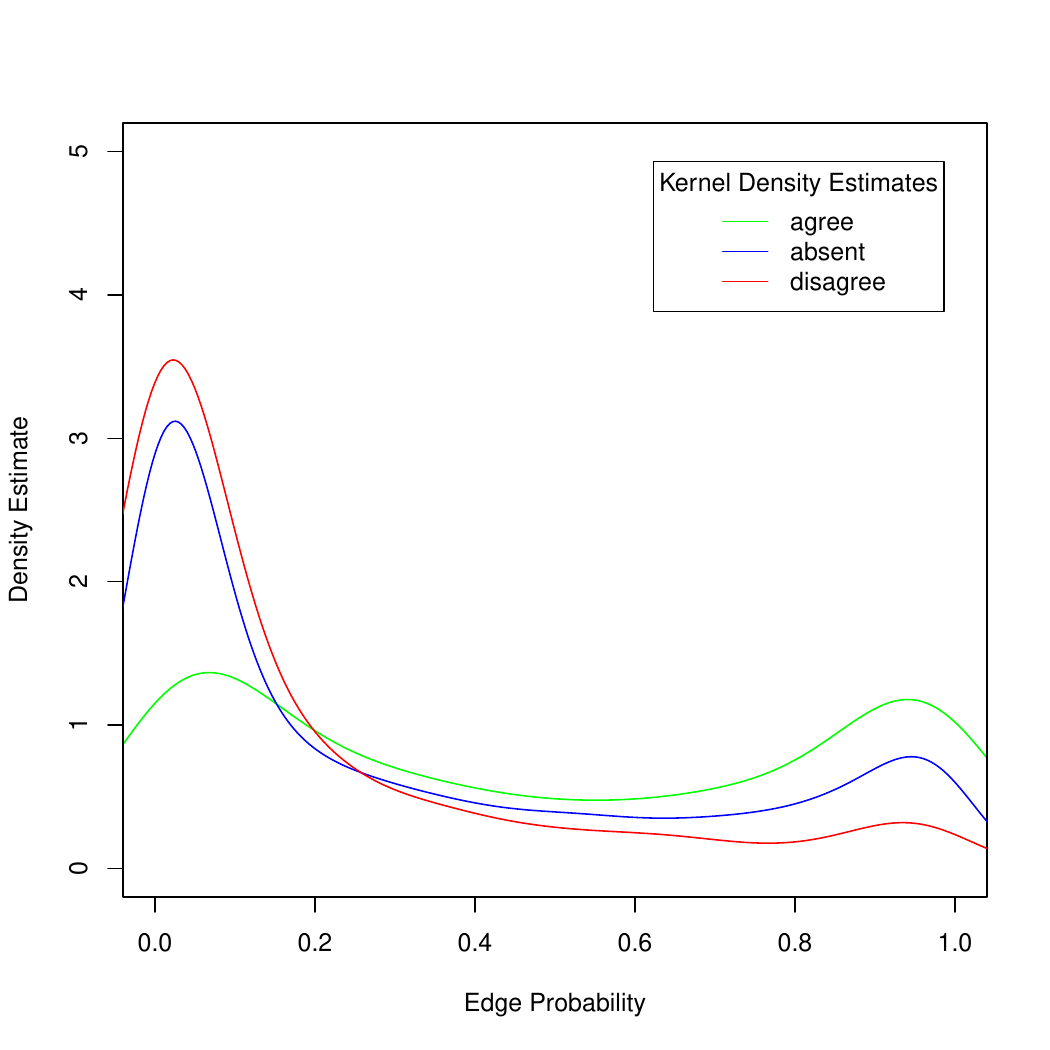} & 
\includegraphics[width=0.3\textwidth]{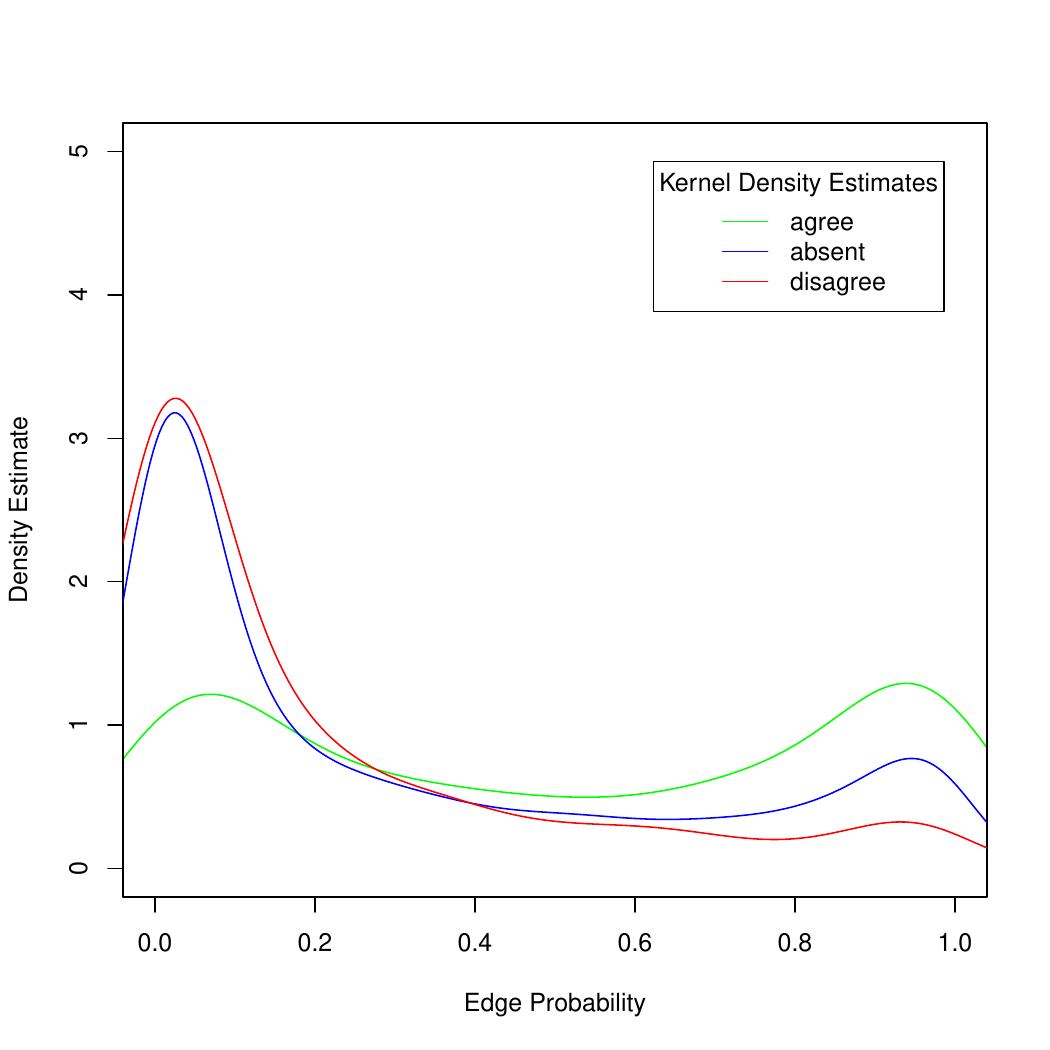} \\
\end{tabular}
\newpage
\begin{tabular}{ccc}
Vote 10 & Vote 11 & Vote 12\\
\includegraphics[width=0.3\textwidth]{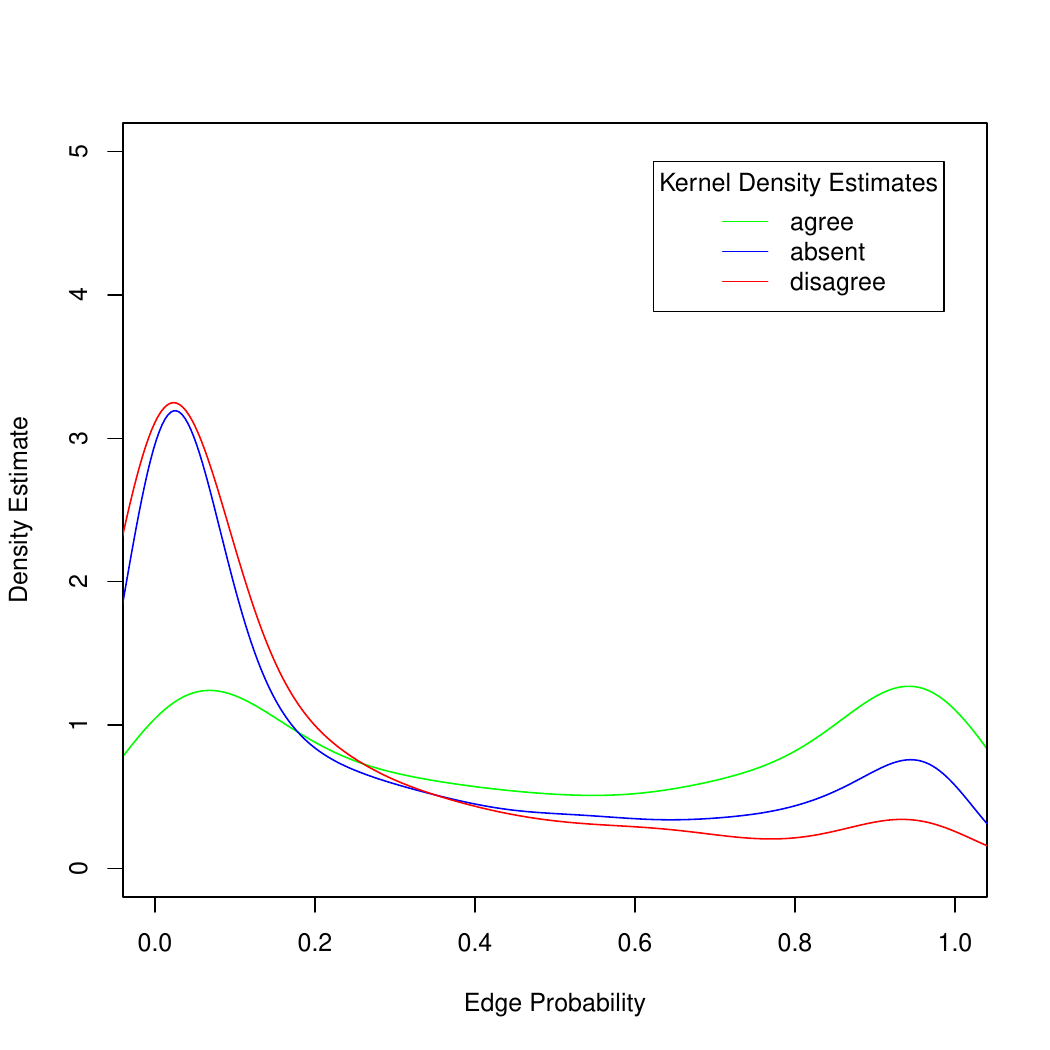} & \includegraphics[width=0.3\textwidth]{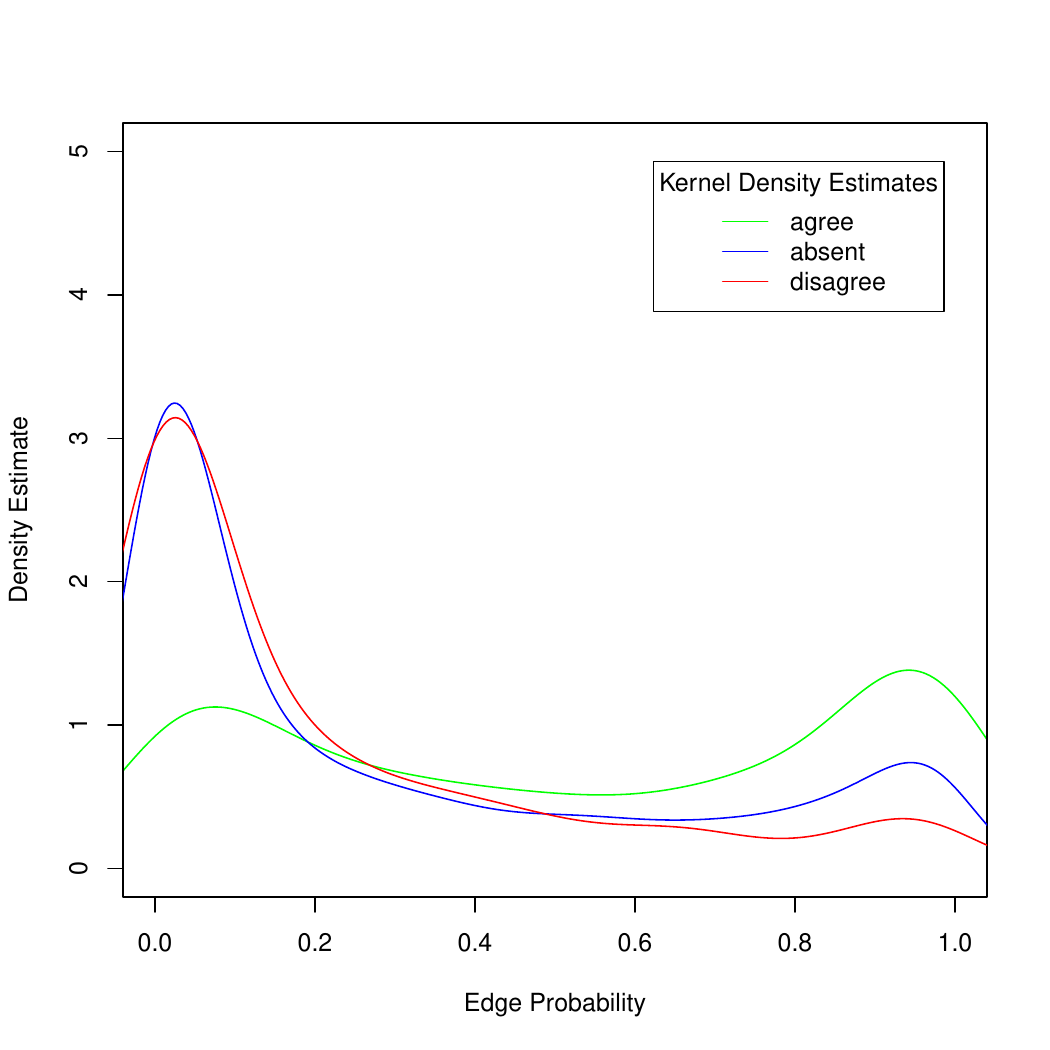} & 
\includegraphics[width=0.3\textwidth]{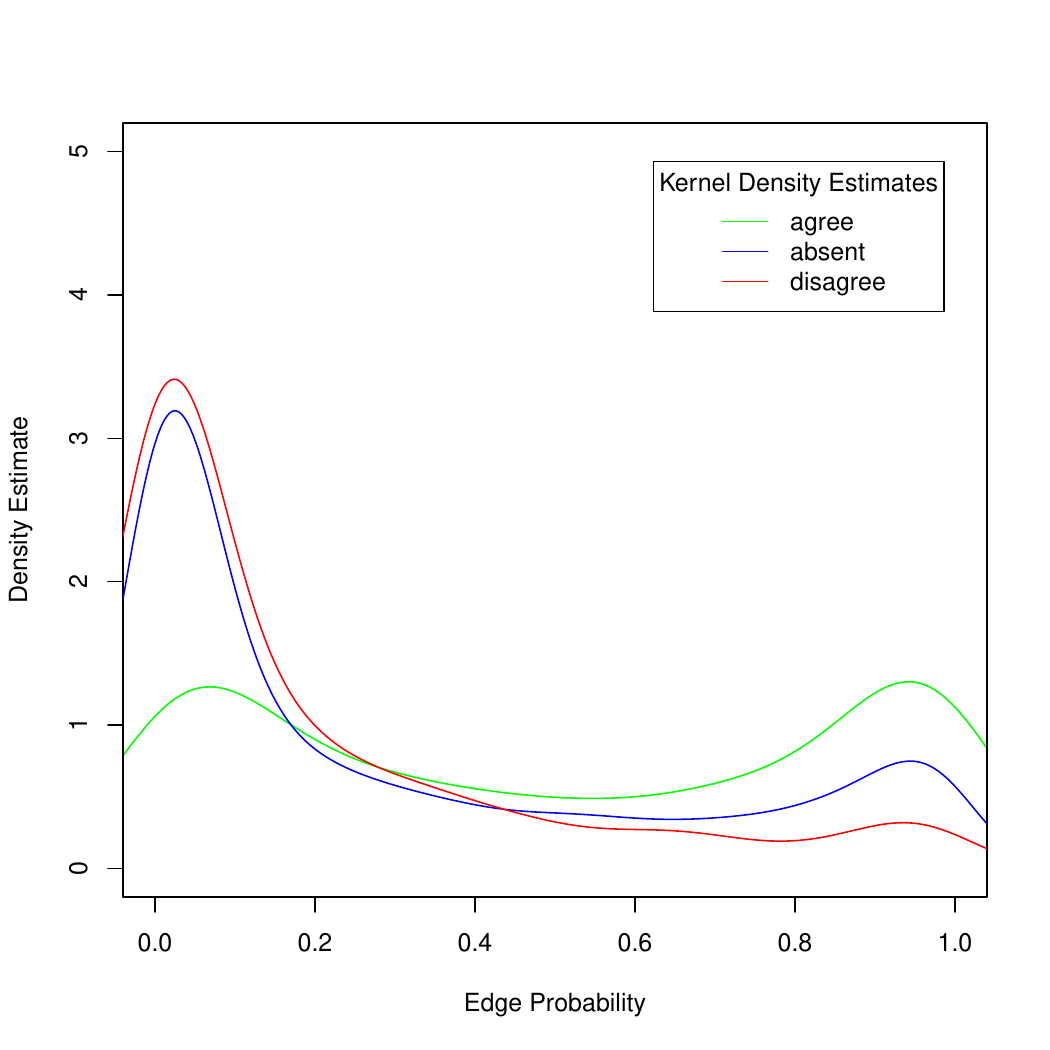} \\
Vote 13 & Vote 14 & Vote 15\\
\includegraphics[width=0.3\textwidth]{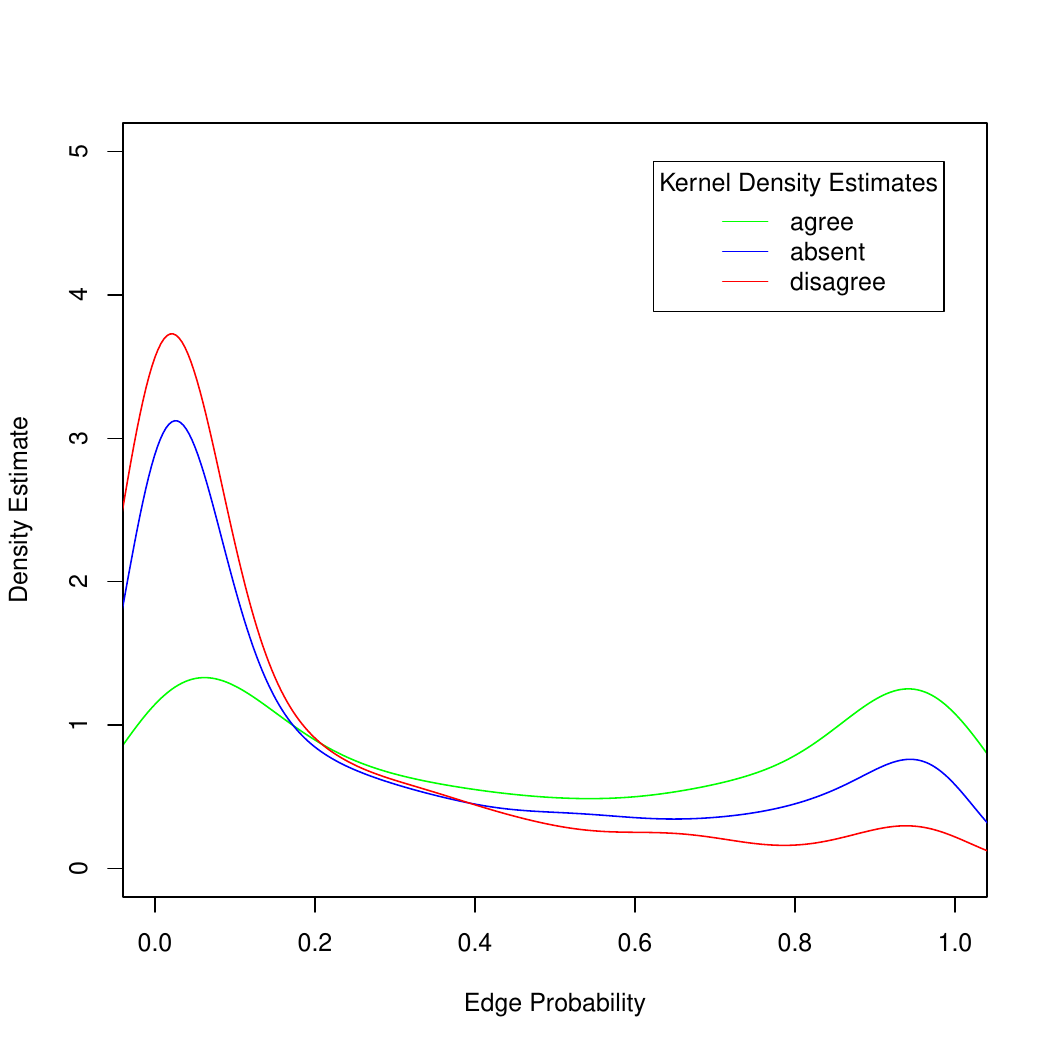} & \includegraphics[width=0.3\textwidth]{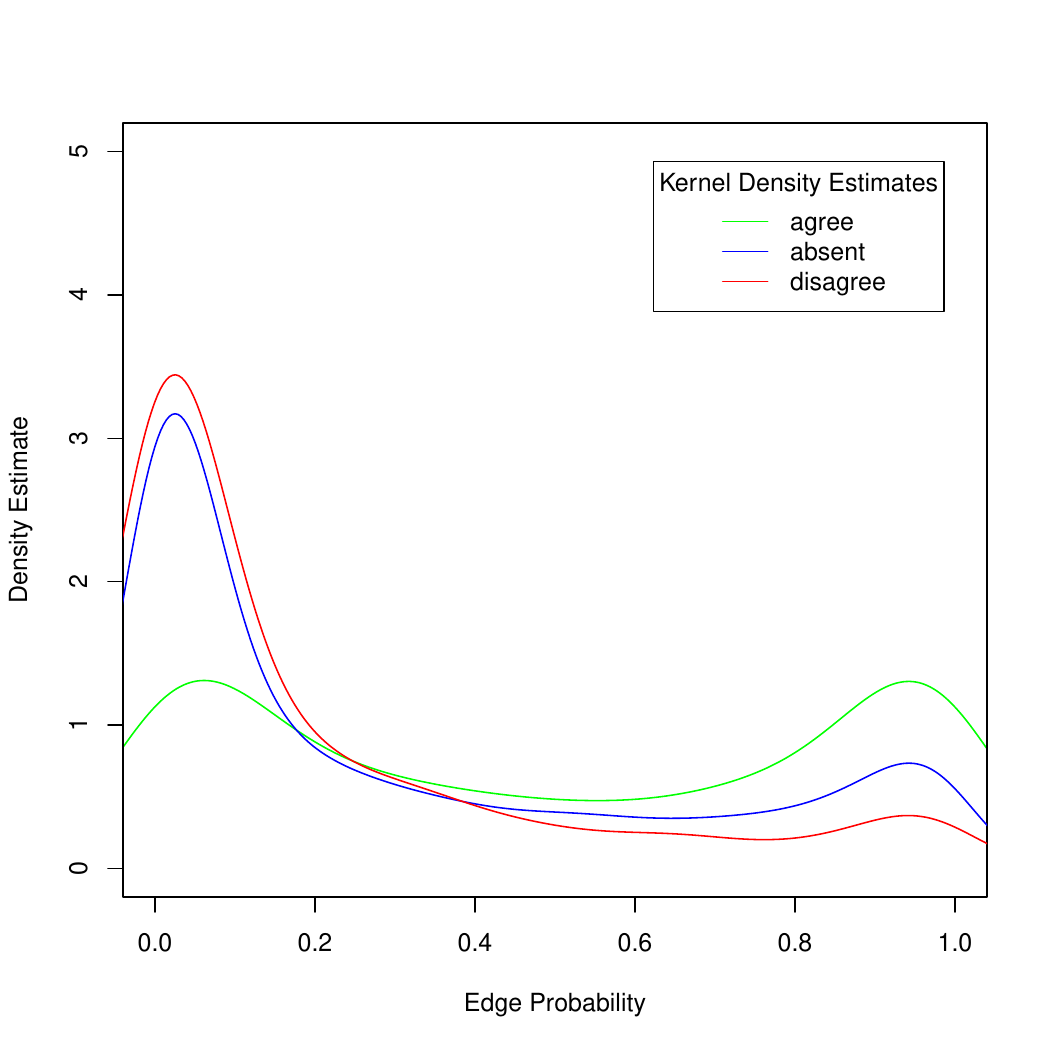} & 
\includegraphics[width=0.3\textwidth]{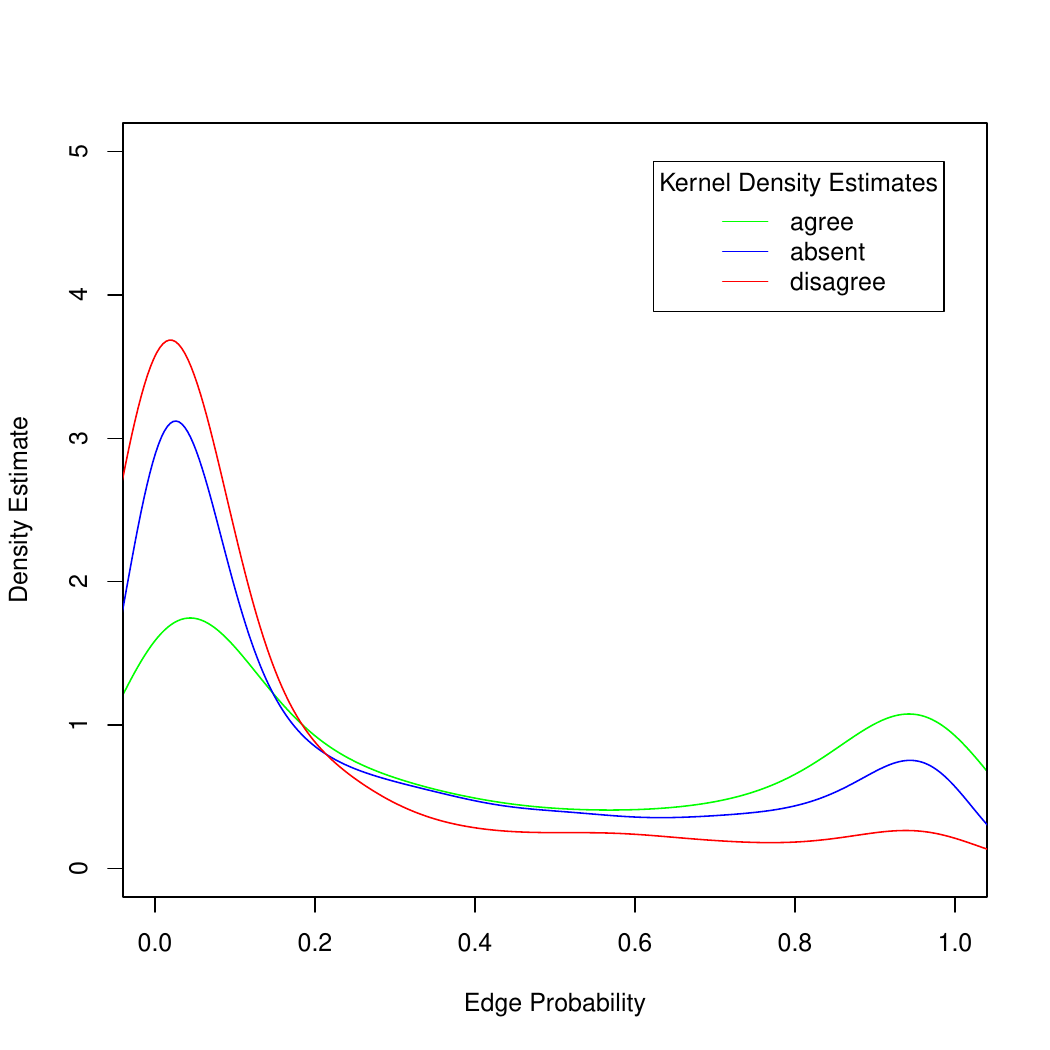} \\
Vote 16 & Vote 17 & Vote 18\\
\includegraphics[width=0.3\textwidth]{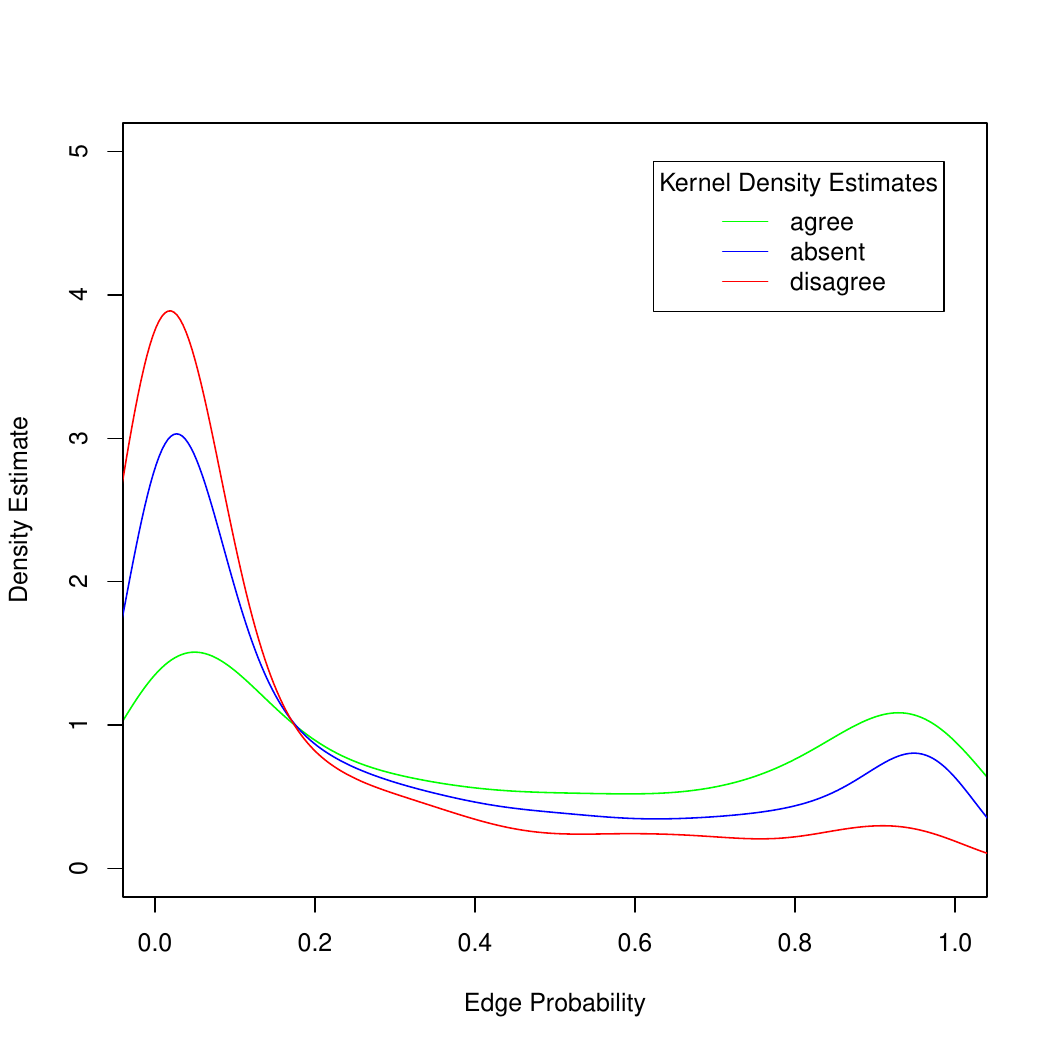} & \includegraphics[width=0.3\textwidth]{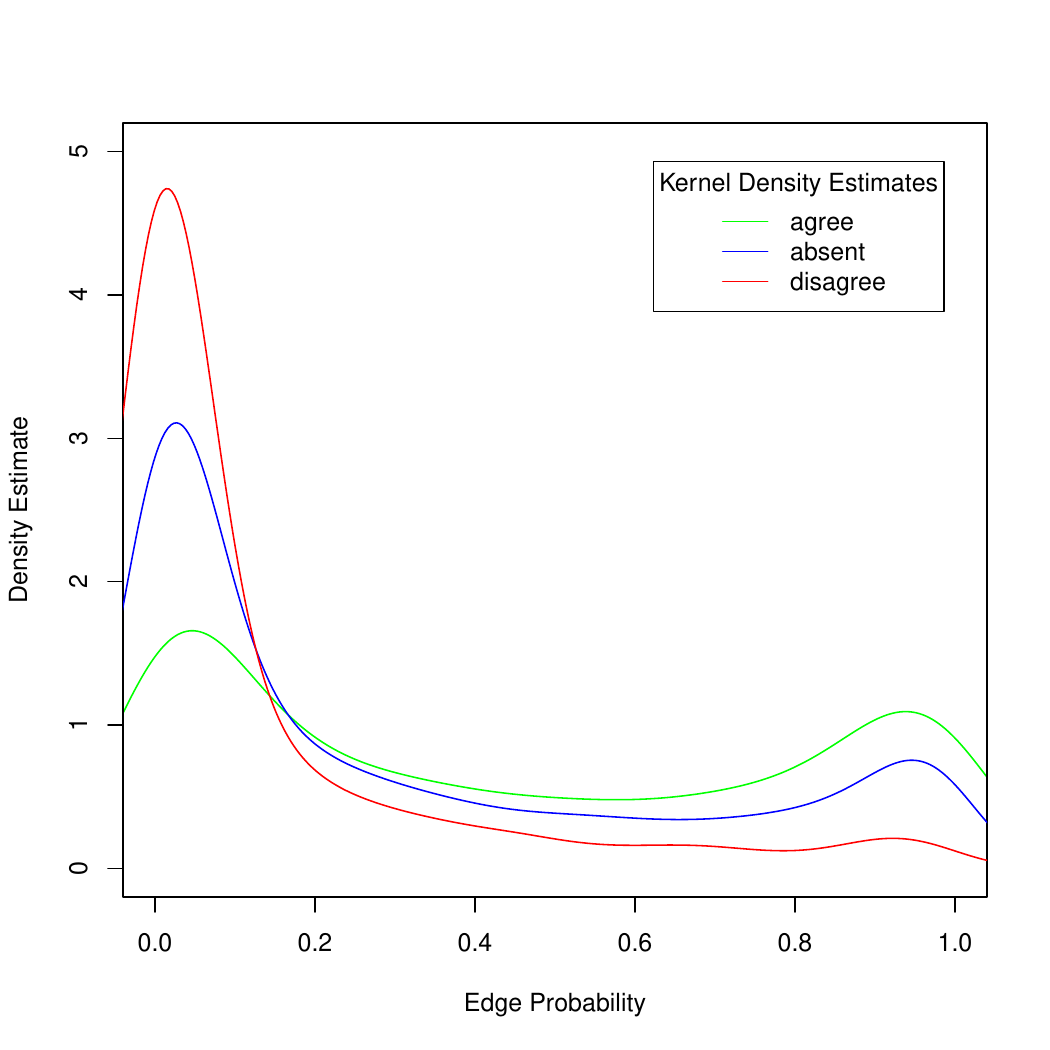} & 
\includegraphics[width=0.3\textwidth]{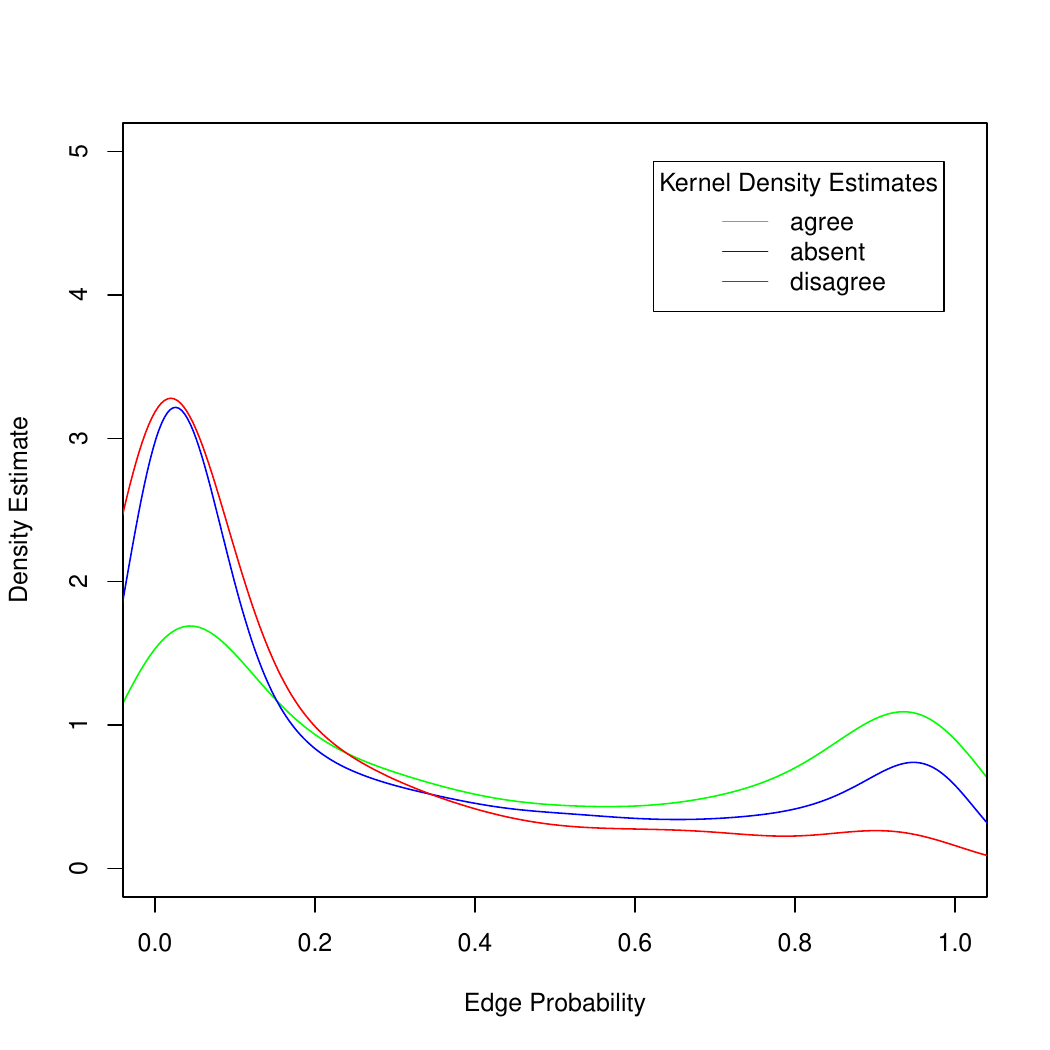} \\
\end{tabular}
\newpage

\begin{tabular}{ccc}
Vote 19 & Vote 20 & Vote 21\\
\includegraphics[width=0.3\textwidth]{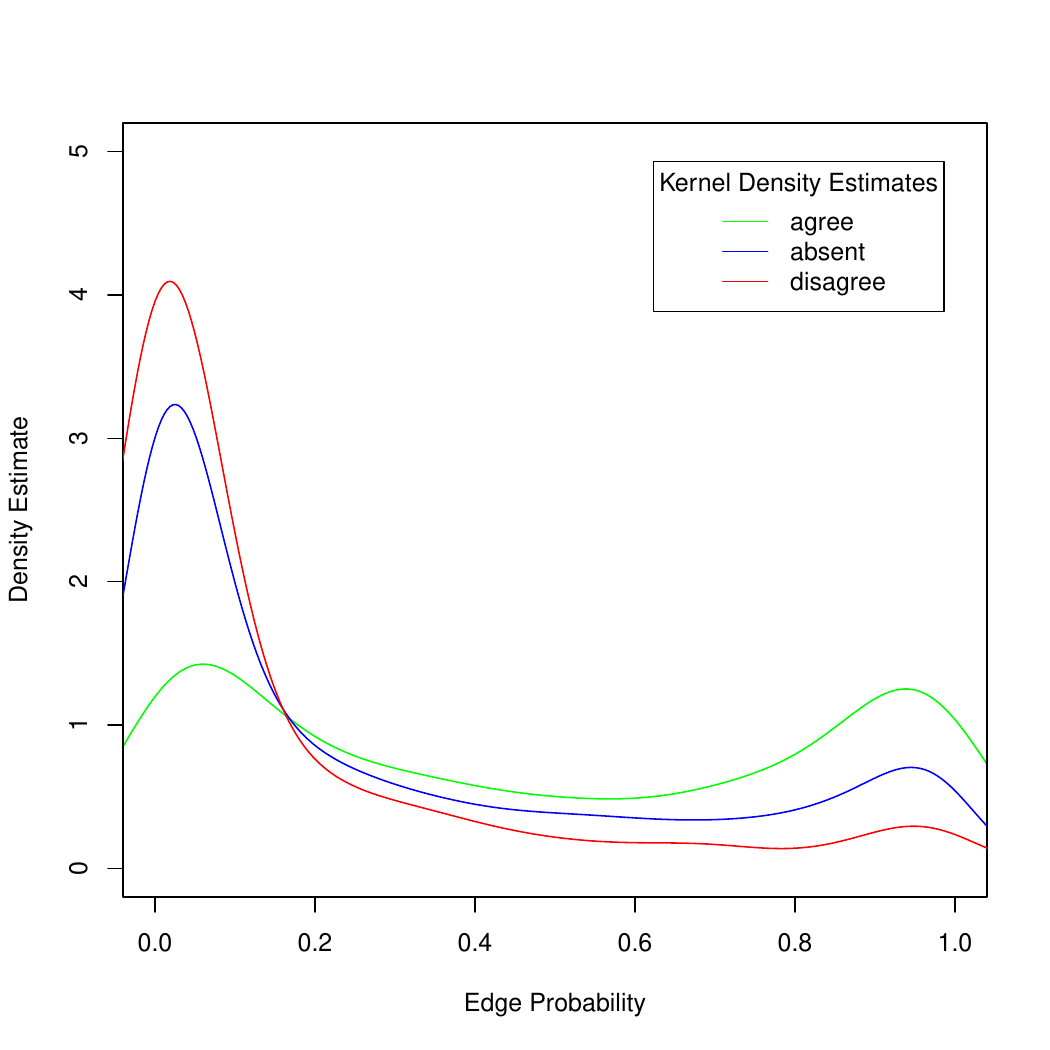} & \includegraphics[width=0.3\textwidth]{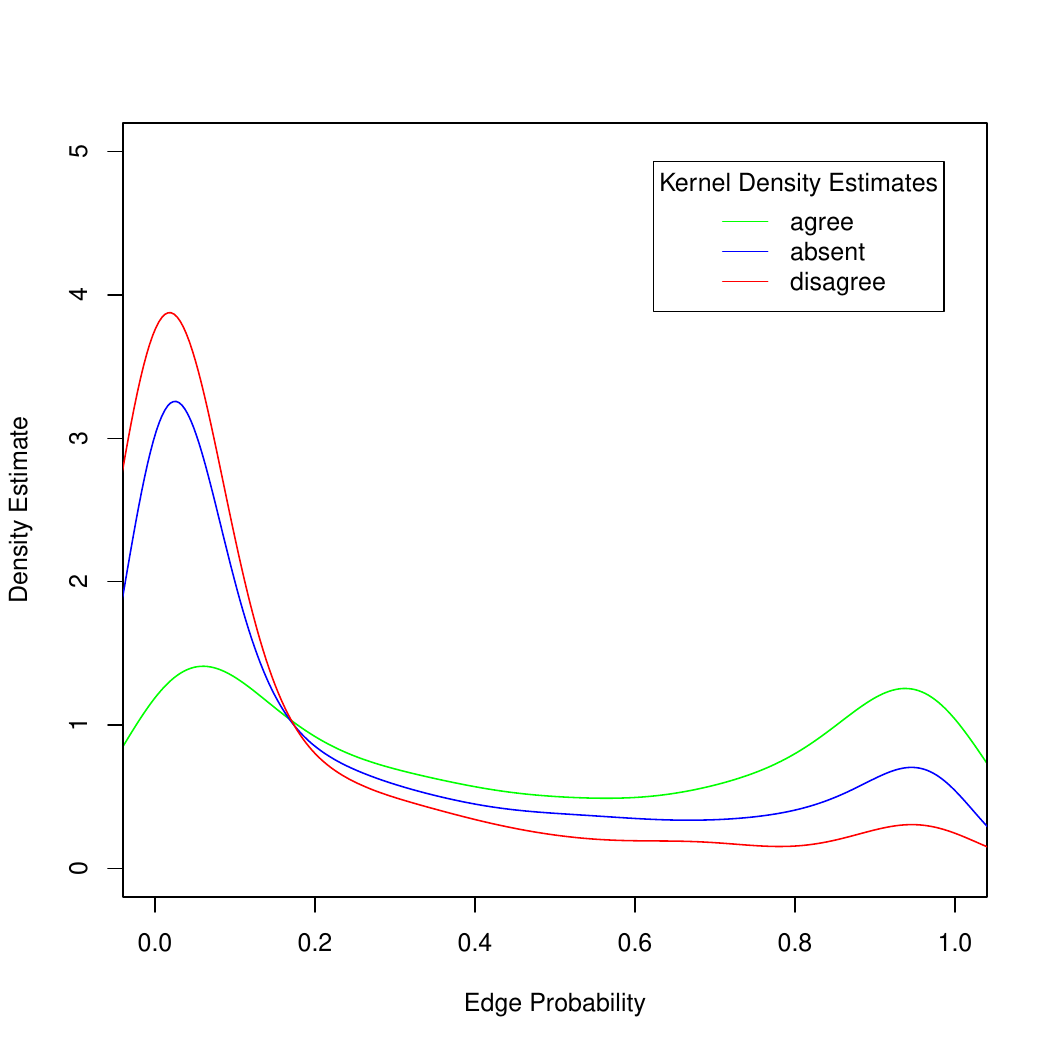} & 
\includegraphics[width=0.3\textwidth]{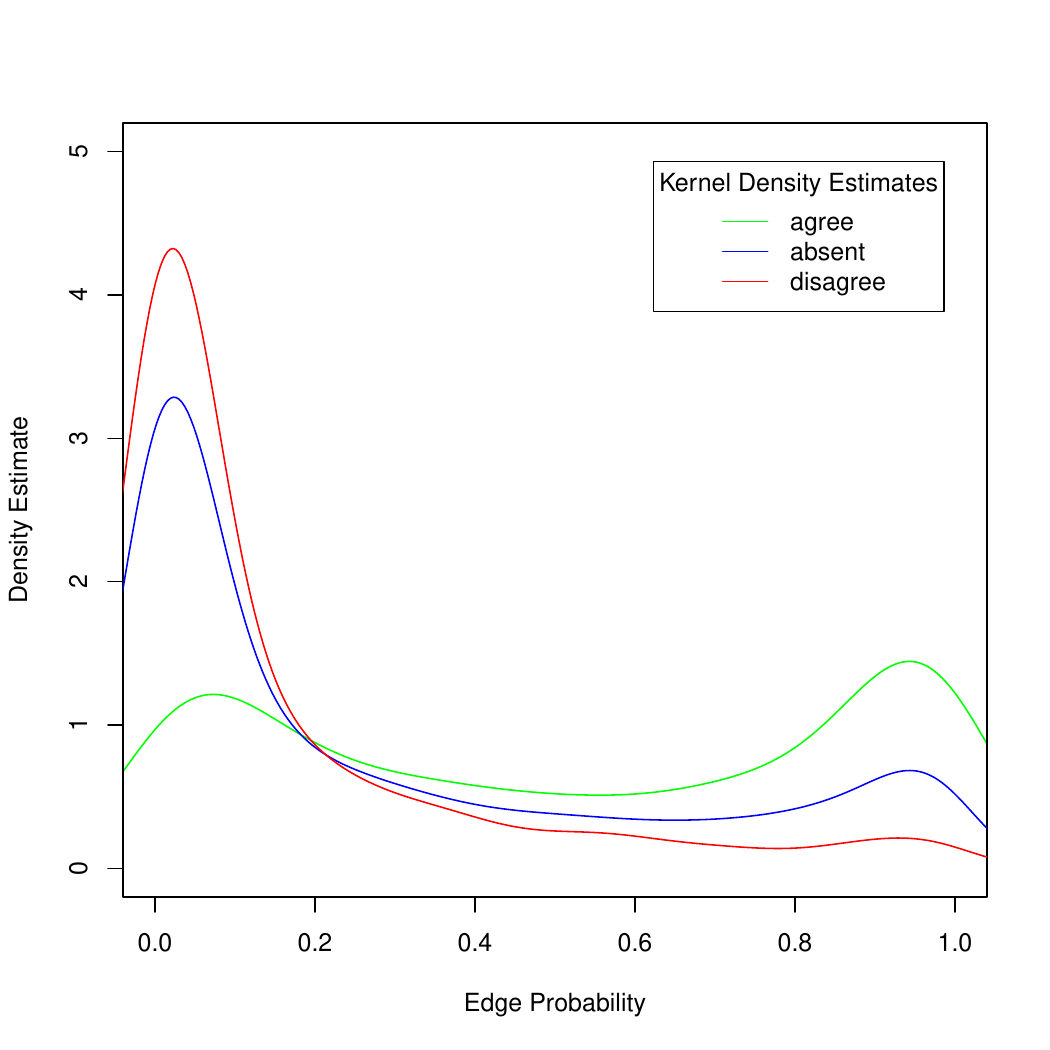} \\
Vote 22 & \\
\includegraphics[width=0.3\textwidth]{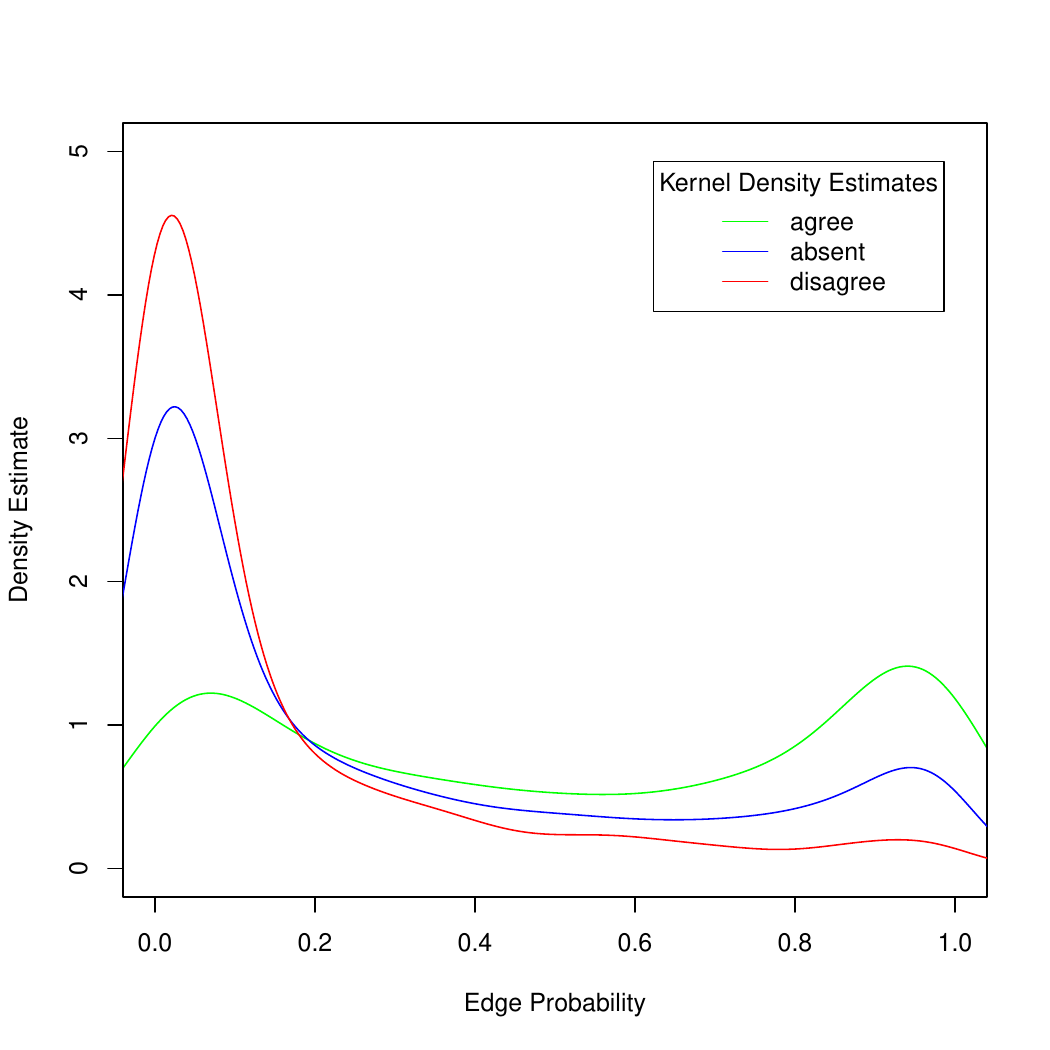}  \\
\end{tabular}

\newpage

\end{document}